\documentclass[aps,prx,amsmath,twocolumn,groupedaddress,letterpaper,floatfix,longbibliography]{revtex4-2}

\usepackage{graphicx,subfigure,epsfig}
 \usepackage{array,multirow}
\usepackage{dcolumn}
\usepackage{amssymb,amsmath,amsfonts,mathrsfs}
\usepackage{array}
\usepackage{times,setspace}
\usepackage{latexsym}
\usepackage{float,flafter,bm,bbm}
\usepackage{epstopdf,color,multirow}
\usepackage[colorlinks,linkcolor=blue,anchorcolor=blue,urlcolor=blue,citecolor=blue]{hyperref}
\usepackage{footnote}
\usepackage{booktabs}
\usepackage{xcolor}

\makeatletter
\usepackage{dcolumn}
\usepackage{bm}
\usepackage{dsfont}
\usepackage{multirow}
\usepackage{color}

\makeatother

\begin{document}
\title{ 
Generalized Peierls substitution for Wannier obstructions: response to  disorder and interactions
}
\author{Shuai A. Chen}
\email{chsh@pks.mpg.de}
\affiliation{Max Planck Institute for the Physics of Complex Systems, N\"{o}thnitzer Stra{\ss}e 38, Dresden 01187, Germany}

\author{Roderich Moessner}
\affiliation{Max Planck Institute for the Physics of Complex Systems, N\"{o}thnitzer Stra{\ss}e 38, Dresden 01187, Germany}

\author{Tai Kai Ng}
\affiliation{Thrust of Advanced Materials, Hong Kong University of Science and Technology (Guangzhou), Guangzhou, China}
\affiliation{Department of Physics, Hong Kong University of Science and Technology,
Clear Water Bay, Hong Kong, China}
\date{\today}

\begin{abstract}
We study the interplay between quantum geometry, interactions, and external fields in complex band systems. When Wannier obstructions preclude a description based solely on atomic-like orbitals,
this complicates the prediction of electromagnetic responses particularly in the presence of disorder and interactions. In this work, we introduce a generalized Peierls substitution framework based on Lagrange multipliers to enforce the constraints of the Wannier obstruction in the band of interest. Thus we obtain effective descriptions of interactions and disorder in the presence of non-trivial quantum geometry of that band. 
We apply our approach to examples including the diamagnetic response in flat-band superconductors and delocalization effects in flat-band metals caused by interactions and disorder.
\end{abstract}

\maketitle

\begin{figure}
\centering 
\includegraphics[scale=0.6]{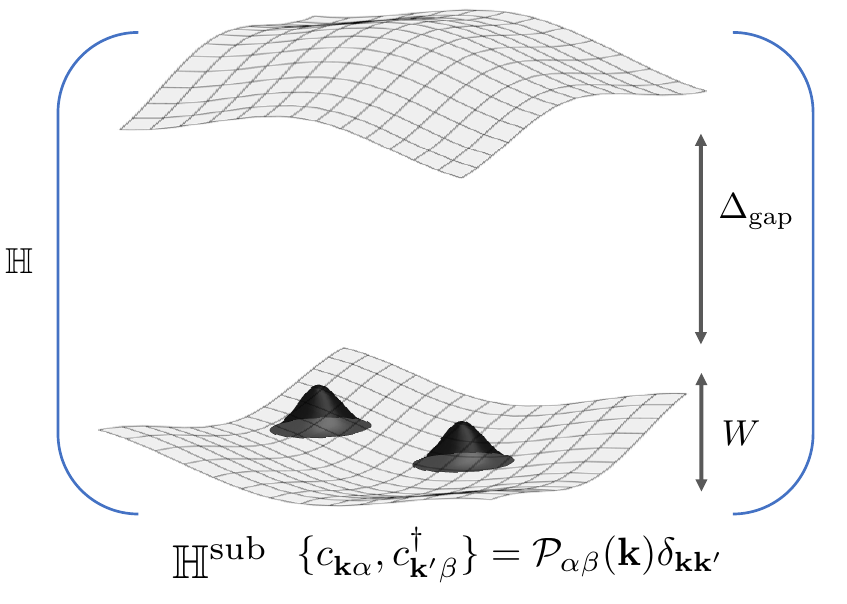} 
\caption{
Schematic illustration of bands of Wannier obstructions. With a large band gap $\Delta_\mathrm{gap}$, the low-energy physics can be captured in the subspace $\mathbb{H}^{\mathrm{sub}}$. Electrons are not independent across different orbitals $\{c_{\mathbf{k}\alpha}^ {},c_{\mathbf{k}^{\prime}\beta}^{\dagger}\}=\mathcal{P}_{\alpha\beta}(\mathbf{k})\delta_{\mathbf{k}\mathbf{k}^{\prime}}$ in the subspace $\mathbb{H}^{\mathrm{sub}}$. The kernel $\mathcal{P}_{\alpha\beta}(\mathbf{k})$ is the band projector under the orbital basis. 
}
\label{Fig:bandstructure} 
\end{figure}

\emph{Introduction.---} 
Recent advances in materials science, specially the emergence of moir\'e materials, challenge our understanding of
correlated band systems of non-trivial quantum geometry (QG)
\cite{2018Natur55643C,2018Natur.556...80C,2011PNAS..10812233B,2018PhRvB..98d5103Y,2018PhRvX...8c1089P,2019SciBu..64..310H,2018PhRvL.121h7001X,2019PhRvB..99l1407R,2018PhRvB..97w5453G,2019PhRvB..99m4515R,2023Natur.614..440T}. 
These systems have isolated bands which are separated from remote bands by a large band gap
$\Delta_{\mathrm{gap}}$
(see Fig.~\ref{Fig:bandstructure}) while the characteristic interaction strengths are much weaker than the band gap. 
Importantly, QG can manifest through Wannier obstructions which preclude a description based solely on atomic-like orbitals. The geometric contributions can introduce geometric contributions such as Berry curvature governing the topological phase \cite{1985AnPhy.160..343K,PhysRevLett.93.206602,RevModPhys.82.1959} and quantum metric encoding the distinguishability of Bloch states \cite{1980CMaPh76289P}. It can reshape the low-energy physics to go beyond knowledge of band dispersion \cite{2018PhRvL.121l6402P,2018PhRvB..98h5435Z,PhysRevLett.128.176403,2019PhRvX...9b1013A,2018PhRvX...8c1089P,2024NatRM...9..481A,2019Natur.574..653L,PhysRevX.9.031049,PhysRevLett.123.237002,PhysRevB.101.060505,PhysRevResearch.2.023237,2020PhRvL124p7002X,PhysRevResearch4013164,PhysRevB.104.115160,PhysRevResearch.4.013209,2021arXiv211100807T,PhysRevLett.127.246403,PhysRevLett.128.176403}. 
This is particularly evident in flat-band superconductors, where the Fermi velocity vanishes, and quantum metric plays a key role in determining superfluid weight and coherence length \cite{2015NatCo68944P,PhysRevB.95.024515,PhysRevA.97.033625,PhysRevB.98.220511,2020arXiv200716205J,PhysRevB.102.201112,PhysRevA.101.053631,PhysRevLett.117.045303,PhysRevLett.127.170404,PhysRevLett.128.087002,PhysRevB.106.014518,PhysRevB.105.L140506,PhysRevB.106.104514,2022arXiv220900007H,2022arXiv220402994H,2016PhRvB..94x5149T,2023PNAS..12017816M,2021PNAS11806744V,2024PhRvL.132b6002C,2023arXiv230805686H,2024arXiv240912254H,PhysRevResearch.6.013256,2024arXiv241023121V,PhysRevLett.131.240001,PhysRevB.109.214518}.

One can construct an effective theory through the band projections into the subspace associated with these isolated bands. 
However, problems can arise when we consider the electromagnetic response. For instance, imposing band projections directly on both the Hamiltonian and the current operator for a Landau level would yield immobile electrons as the kinetic energy in the Landau level is quenched. But in actuality, in this problem, it is the projected Coulomb interaction in the lowest Landau level which can control the highly nontrivial charge dynamics~\cite{1993IJMPB...7.4389M,1994MPLB....8.1065R}. 
In lattice systems, electromagnetic fields are incorporated via the Peierls substitution through attaching phase factors to hopping terms. This method approximates continuum minimal coupling by assuming exponentially localized Wannier functions, which allows truncating the hopping to nearest neighbors. However, its application is challenged on Wannier-obstructed bands where no complete set of maximally localized Wannier functions exists. The obstructions, exemplified by topological bands, can force tight-binding models to include long-range hopping terms which can produce non-local phase factors and thus a non-local current operator.

This raises the question: can we establish a theoretical framework self-consistently within the subspace $\mathbb{H}^{\mathrm{sub}}$? In particular, can we generalize the Peierls substitution to topological bands of Wannier functions to obtain the current operator in the presence of disorder or interactions? 
Some progress on this has been made in identifying a multiplicative phase factor from the effective projected interaction terms \cite{2023PNAS..12017816M,2021PNAS11806744V} in response to an electromagnetic field. 
However, these approaches lack a general method for deriving the current operator in the subspace $\mathbb{H}^{\mathrm{sub}}$ as they only consider specific cases.

In this paper, we generalize the Peierls substitution framework to resolve the response to disorder and interactions in bands of Wannier obstructions. We first enforce the band projection by a set of Grassmann-valued Lagrange multipliers through $L_\mathrm{\lambda}$ in Eq.~(\ref{eq:Llambda})
onto the subspace $\mathbb{H}^{\mathrm{sub}}$. 
The generalization of Peierls substitution is achieved by identifying the gauge coupling through $L_\mathrm{\lambda}$. We apply our approach to two examples: the diamagnetic response in flat-band superconductors and the delocalization effect in flat-band metals, where the currents from the interaction and disorder terms dominate the low-energy dynamics.

\emph{Model setup.---} We consider a Hamiltonian $H=H_{0}+H_{1}$, where the free part $H_0$ describes a band structure which possesses isolated bands with a band gap $\Delta_\mathrm{gap}$ as depicted in Fig.~\ref{Fig:bandstructure}.
The term $H_1$ includes additional contributions that cannot be represented within the free part $H_0$, such as  interaction and disorder terms. 
The energy scales of the additional terms in $H_1$ are much weaker than the band gap $\Delta_\mathrm{gap}$.
The tight-binding Hamiltonian $H_0$ contains $N_o$ orbitals,
\begin{equation}
H_{0}=\sum_{\alpha,\beta=1
\cdots N_o}\sum_{\mathbf{k}}h^{}_{\alpha\beta}(\mathbf{k})c_{\mathbf{k}\alpha}^{\dagger}c_{\mathbf{k}\beta}^ {},
\end{equation}
where the operator $c_{\mathbf{k}\alpha}$ ($c_{\mathbf{k}\alpha}^{\dagger}$)
annihilates (creates) a fermion at an orbital $\alpha$ of the crystal
momentum $\mathbf{k}$ and satisfies the anticommutators
$\{c_{\mathbf{k}\alpha},c_{\mathbf{k}^{\prime}\beta}^{\dagger}\}=\delta_{\mathbf{k}\mathbf{k}^{\prime}}\delta_{\alpha\beta}$.
For simplicity, we assume only one band at the Fermi energy
with the wave function $u_{\mathbf{k}\alpha}$: $\sum_{\beta}h_{\alpha\beta}(\mathbf{k})u_{\mathbf{k}\beta}=\epsilon_{\mathbf{k}}u_{\mathbf{k}\alpha}$,
while these states span the subspace $\mathbb{H}^{\mathrm{sub}}$.
We derive a low-energy
theory in the subspace $\mathbb{H}^{\mathrm{sub}}$ of
the partially filled band for $\Delta_{\mathrm{gap}}\rightarrow\infty$.

\emph{Band projection and Lagrange multipliers.---}
As depicted in Fig.~\ref{Fig:bandstructure}, 
we introduce a band projection
\begin{equation}
c_{\mathbf{k}\alpha}^{\dagger}\rightarrow u_{\mathbf{k}\alpha}f_{\mathbf{k}}^{\dagger},\label{eq:proj}
\end{equation}
with $f_{\mathbf{k}}^\dagger$ 
satisfying the anti-commutation $\{f_{\mathbf{k}},f_{\mathbf{k}^{\prime}}^{\dagger}\}=\delta_{\mathbf{k}\mathbf{k}^{\prime}}$ and creating a fermion with wave function $u_{\mathbf k\alpha}$ in the isolated band. 
The projection is unique up to an arbitrary phase
of wave functions: $u_{\mathbf{k}\alpha}\rightarrow u_{\mathbf{k}\alpha}e^{i\phi(\mathbf{k})}$.
Within the sub-Hilbert space $\mathbb H^\mathrm{sub}$, the projected $c_{\mathbf{k}\alpha}$-fermions obey unconventional anti-commutators,
\begin{equation}
\!\!\!\!\{c_{\mathbf{k}\alpha}^{},c_{\mathbf{k}^{\prime}\beta}^{\dagger}\}\!\!=\!\mathcal{P}_{\alpha\beta}(\mathbf{k})\delta_{\mathbf{k}\mathbf{k}^{\prime}},\{c_{\mathbf{k}\alpha},c_{\mathbf{k}^{\prime}\beta}\}\!\!=\!\!\{c_{\mathbf{k}\alpha}^{\dagger},c_{\mathbf{k}^{\prime}\beta}^{\dagger}\!\}\!=0.\label{eq:uncon}
\end{equation}
The kernel $\mathcal{P}_{\alpha\beta}(\mathbf{k})\equiv u_{\mathbf{k}\alpha}^{*}u_{\mathbf{k}\beta}^{}$ is the band projector under the orbital basis. It is not a constant function of  momentum $\mathbf k$ when the band possesses the Wannier obstruction \cite{PhysRevB.56.12847}.
One  profound consequence
is a generalized Girvin-MacDonald-Platzman (GMP) algebra \cite{PhysRevLett.54.581,PhysRevB.33.2481,2024arXiv241000969W}
for the projected density operator $\rho(\mathbf{q})\equiv\sum_{\mathbf{k}\alpha}c_{\mathbf{k}\alpha}^{\dagger}c_{\mathbf{k}+\mathbf{q}\alpha}^{}$.

We highlight two challenges arising from the projection in Eq.~(\ref{eq:proj}).
First, we cannot independently vary $c_{\mathbf{k}\alpha}$ for
different orbitals when e.g. deriving the dynamics through
the variational method. Second, the
projection may not commute with the action of an external perturbation. For example,
the presence of the electromagnetic field can change the eigenfunction of $H_0$, and the projection in Eq.~\eqref{eq:proj} should change accordingly.

To enforce Eq.~\eqref{eq:uncon}, we introduce the complex Grassmann valued Lagrange multipliers $\lambda_{\mathbf{k}\alpha}^{},\lambda_{\mathbf{k}\alpha}^{\dagger}$ to realize the projection in Eq.~(\ref{eq:proj}), 
\begin{equation}
L_{\lambda}=\sum_{\mathbf{k}\alpha}\lambda_{\mathbf{k}\alpha}^{\dagger}\left(c_{\mathbf{k}\alpha}-u_{\mathbf{k}\alpha}^{*}f_{\mathbf{k}}\right)+h.c..\label{eq:Llambda}
\end{equation}
(For a different scheme, see Ref.~\cite{1993IJMPB...7.4389M}.) Eq.~\eqref{eq:Llambda} implies a $U(1)$ symmetry in the subspace $\mathbb{H}^{\mathrm{sub}}$. Concretely, with the Wannier function $\varphi_{\alpha}(\mathbf{r})=\sum_{\mathbf{\mathbf{k}\in \mathrm{BZ}}}e^{i\mathbf{k}\cdot\mathbf{r}}u_{\mathbf{k}\alpha}$ of orbital $\alpha$ centered at the original point, we have in real space 
\begin{align}
L_{\lambda}=\sum_{\mathbf{r}\alpha}\lambda_{\mathbf{r}\alpha}^{\dagger}[c_{\mathbf{r}\alpha}-\sum_{\mathbf{r}^\prime}\varphi_{\alpha}^{*}(\mathbf{r}-\mathbf{r}^{\prime})f_{\mathbf{r}^{\prime}}]+h.c.,
\label{Ll_r}
\end{align}
and then the $U(1)$ symmetry manifests as
\begin{equation}
\begin{split}
&\lambda_{\mathbf{r}\alpha}\rightarrow\lambda_{\mathbf{r}\alpha}e^{-i\theta},c_{\mathbf{r}\alpha}\rightarrow c_{\mathbf{r}\alpha}e^{-i\theta},f_{\mathbf{r}}\rightarrow f_{\mathbf{r}}e^{-i\theta}, \\
&\lambda^\dagger_{\mathbf{r}\alpha}\rightarrow\lambda^\dagger_{\mathbf{r}\alpha}e^{i\theta},c^\dagger_{\mathbf{r}\alpha}\rightarrow c^\dagger_{\mathbf{r}\alpha}e^{i\theta},f^\dagger_{\mathbf{r}}\rightarrow f^\dagger_{\mathbf{r}}e^{i\theta},
\end{split}
\label{eq:U(1)transf}
\end{equation}
where $c_{\mathbf{r}\alpha}=\sum_{\mathbf{k}\in \mathrm{BZ}}c_{\mathbf{k}\alpha}e^{i\mathbf{k}\cdot\mathbf{r}}$
 and similar expressions
hold for $\lambda_{\mathbf{r}\alpha}$ and $f_{\mathbf{r}}$. 
The $U(1)$ symmetry in Eq.~\eqref{eq:U(1)transf} indicates a Noether current. 
The formula for multiple bands at the Fermi energy can be expressed similarly \cite{NOTEMB}.

We can summarize the effective theory within the subspace $H^\mathrm{sub}$ 
with the partition function $Z=\int D\lambda D\lambda^{\dagger}DcDc^{\dagger}Df_{\mathbf{}}Df^{\dagger}e^{i\int dtL}$ with 
\begin{equation}
L=\sum_{\mathbf{k}\alpha}c_{\mathbf{k}\alpha}^{\dagger}i\partial_{t}c_{\mathbf{k}\alpha}^ {}-H_{0}-H_{1}-L_{\lambda}.
\label{eq:Leff}
\end{equation}
After integrating out the Lagrange multipliers, we recover the constraint $c_{\mathbf{k}\alpha}=u_{\mathbf{k}\alpha}^{*}f_{\mathbf{k}}$. 
The constraint needs to be varied when we apply external perturbations. In the following, we  derive the current response from both Euler-Lagrange equations and the Peierls substitution based on the effective Lagrangian in Eq.~\eqref{eq:Leff}.

\emph{Euler-Lagrange equation and Noether current.---}
The $U(1)$ transformation in Eq.~(\ref{eq:U(1)transf}) leads to a conserved Noether current. 
We derive the Euler-Lagrange (EL) equations first, and then proceed to derive the continuity equations.
For simplicity, we assume that the terms in $H_1$, such as the Hubbard interaction, commute with the local density operators in the full Hilbert space $\mathbb H$ and therefore do not contribute to the current operator in $\mathbb H$.  We will identify two contributions to the Noether current, namely from band dispersion and interactions in the subspace $\mathbb H^\mathrm{sub}$. The EL equation can be obtained from the functional derivative $\frac{\delta S}{\delta\phi}=0,\quad\phi\in\{c,c^{\dagger},\lambda,\lambda^{\dagger},f,f^{\dagger}\}$
with the action $S=\int dtL$. Above all, we can figure out the Lagrange multipliers $\lambda_{\mathbf{k}\alpha}=\sum_{\beta}\left[\mathcal{P}_{\alpha\beta}(\mathbf{k})-\delta_{\alpha\beta}\right]\frac{\delta H_{1}}{\delta c_{\mathbf{k}\beta}^{\dagger}}$, 
which does not vanish when both interaction and non-constant kernel $\mathcal P_{\alpha\beta}$ are present.
Then we can reach the EL equation with  details in Supplementary Materials (SM) \cite{SM},
\begin{equation}
    i\partial_{t}c_{\mathbf{k}\alpha}  =\sum_{\beta}h_{\alpha\beta}(\mathbf{k})c_{\mathbf{k}\beta}+\sum_{\beta}\mathcal{P}_{\alpha\beta}(\mathbf{k})\frac{\delta H_{1}}{\delta c_{\mathbf{k}\beta}^{\dagger}},
    \label{eq:dck}
\end{equation} 
while the dynamics of $i\partial_{t}c_{\mathbf{k}\alpha}^{\dagger}$ is obtained by taking the Hermitian conjugate of Eq.~\eqref{eq:dck}.
The kernel $\mathcal{M}_{\alpha\beta}(\mathbf{k})$ appears in
the interaction term, which can contribute to the continuity equation,
\begin{align}
\!\!-  i\partial_{t}\rho(\mathbf{q})=&\sum_{\mathbf{k}}(\epsilon_{\mathbf{k}}-\epsilon_{\mathbf{k}+\mathbf{q}})f_{\mathbf{k}+\mathbf{q}}^{\dagger}f_{\mathbf{k}}-\sum_{\mathbf{k}\alpha\beta}[c_{\mathbf{k}+\mathbf{q}\alpha}^{\dagger}\mathcal{P}_{\alpha\beta}(\mathbf{k})\frac{\delta H_{1}}{\delta c_{\mathbf{k}\beta}^{\dagger}} \nonumber 
\\& -\frac{\delta H_{1}}{\delta c_{\mathbf{k}+\mathbf{q}\alpha}}\mathcal{P}_{\alpha\beta}(\mathbf{k}+\mathbf{q})c_{\mathbf{k}\beta}],
\label{eq:continuity}
\end{align}
Here $\rho(\mathbf{q})=\sum_{\mathbf{k}}\langle u_{\mathbf{k}+\mathbf{q}}\vert u_{\mathbf{k}}\rangle f_{\mathbf{k}}^{\dagger}f_{\mathbf{k}+\mathbf{q}}^{}$
is the projected density operator. There are no terms from $H_1$ in Eq.~\eqref{eq:continuity} when the kernel $\mathcal{P}_{\alpha\beta}(\mathbf{k})$ is a constant function of  momentum $\mathbf k$. 
The Noether current $\mathbf{j}_{\mathrm{\mathbf{q}}}$
can be obtained by expanding Eq.~(\ref{eq:continuity}) to the form $\partial_{t}\rho(\mathbf{q})=-i\mathbf{q}\cdot\mathbf{j}_{\mathrm{\mathbf{q}}}$.

From the continuity equation in Eq.~\eqref{eq:continuity}, we can determine two contributions of currents: band dispersion and interactions. 
The current that depends on band dispersion can be obtained by directly projecting the original current operator. 
In contrast, the interaction-driven current is not achievable through this method, and we call it the QG current. 
What kind of $H_1$ can contribute to a non-vanishing QG current? We require that $ H_1$ cannot be expressed in the form of the free Hamiltonian $H_0$. Specifically, it should not take the form of terms like $
\sum_{\mathbf{k} \alpha \beta} V_{\alpha \beta}(\mathbf{k}) c_{\mathbf{k} \alpha}^{\dagger} c_{\mathbf{k} \beta}^{}
$ or $
\sum_{\mathbf{k} \alpha \beta} V_{\alpha \beta}(\mathbf{k}) c_{\mathbf{k} \alpha}^{\dagger} c_{-\mathbf{k} \beta}^{\dagger} + \text{h.c.}$.
We will show that there are QG currents from both the interaction and disorder terms in the examples provided. 
The Noether current is nonlocal as seen by expanding $L_{\lambda}$ in Eq.~\eqref{Ll_r} in real space, 
$L_\lambda =\sum_{\mathbf r\alpha}\{\lambda_{\mathbf r\alpha}^\dagger c_{\mathbf r \alpha }-  \sum_{\mathbf r^\prime n} \frac{1}{n!} [(\mathbf r^\prime- \mathbf r )\cdot \nabla_{\mathbf r}]^n \lambda^\dagger_{\mathbf{r}\alpha}\varphi_\alpha^*(\mathbf r^\prime - \mathbf r) f_\mathbf r\} +h.c.$: 
the Noether current can involve high derivatives
of fields $f$ and $f^\dagger$~\cite{1993IJMPB...7.4389M}.

\emph{Generalized Peierls substitution.---} 
We now generalize the Peierls substitution on bands of Wannier obstructions to derive the current operators. 
In a lattice model, the Peierls substitution is an approximation of the continuum minimal coupling in the response to a slowly varying gauge potential $A_{\mathbf{q}}(t)$ by attaching  $U(1)$ phase factors to hopping terms. 
For Wannier-obstructed bands, the effective Lagrangian in Eq.~\eqref{eq:Leff} contains $L_\lambda$ which realizes the band projection. 
The term $L_{\lambda}$ is invariant under global $U(1)$ transformations, but not under local $U(1)$ transformations. Thus, the $U(1)$ symmetry in Eq.~\eqref{eq:U(1)transf} operates by
involving the $f_{\mathbf{k}\alpha}$,
$c_{\mathbf{k}\alpha}$, $\lambda_{\mathbf{k}\alpha}$ and their conjugates.

According to the Peierls substitution, we can introduce the gauge coupling to the kinetic term $H_{0}[A]$. For a slowly varying gauge potential, we can 
keep the leading order of $A$,
\begin{align}
\!\!H_{0}[A] & =\sum_{\mathbf{k}\alpha\beta}[h_{\alpha\beta}(\mathbf{k})c_{\mathbf{k}\alpha}^{\dagger}c_{\mathbf{k}\beta}^ {}-\sum_{\mathbf{q}\mu}A_{\mathbf{q}}^{\mu}\partial_{\mu}h_{\alpha\beta}(\mathbf{k}+\frac{\mathbf{q}}{2})c_{\mathbf{k}\alpha}^{\dagger}c_{\mathbf{k}+\mathbf{q}\beta}^ {}\nonumber \\
 & +\frac{1}{2}\sum_{\mathbf{q}\mu\nu}A_{\mathbf{q}}^{\mu}A_{-\mathbf{q}}^{\nu}\partial_{\mu}\partial_{\nu}h_{\alpha\beta}(\mathbf{k})c_{\mathbf{k}\alpha}^{\dagger}c_{\mathbf{k}\beta}^ {}],\label{eq:H0A}
\end{align}
with the abbreviation $\partial_{\mu}\equiv\partial/\partial k_{\mu}$.
The terms linear and quadratic in $A$ represent the paramagnetic and
diamagnetic response, respectively. The key step in the generalization
is to introduce a gauge coupling to $L_{\lambda}$, which gives
rise to 
\begin{align}
L_{\lambda}[A]= & L_{\lambda}+L_{\lambda}^{1}+L_{\lambda}^{2},\label{eq:LlambdaA}
\end{align}
with 
\begin{align}
L_{\lambda}^{1} & =\sum_{\mathbf{k}\mathbf{q}\alpha\mu}\lambda_{\mathbf{k}\alpha}^{\dagger}A_{\mathbf{q}}^{\mu}\partial_{\mu}u_{\mathbf{k}+\frac{\mathbf{q}}{2}\alpha}^{*}f_{\mathbf{k}+\mathbf{q}}+h.c.,\\
L_{\lambda}^{2} & =\frac{1}{2}\sum_{\mathbf{k}\mathbf{q}\alpha\mu\nu}\lambda_{\mathbf{k}\alpha}^{\dagger}A_{\mathbf{q}}^{\mu}A_{\mathbf{-q}}^{\nu}\partial_{\mu}\partial_{\nu}u_{\mathbf{k}\alpha}^{*}f_{\mathbf{k}}+h.c..
\end{align}
It is essential to keep both terms $L_{\lambda}^{1}$ and $L_{\lambda}^{2}$
to obey the $U(1)$ gauge symmetry.
Therefore, the Lagrangian $L[A]$ after the Peierls substitution has
the form 
\begin{equation}
L[A]=\sum_{\mathbf{k}\alpha}c_{\mathbf{k}\alpha}^{\dagger}i\partial_{t}c_{\mathbf{k}\alpha}^{}-H_{0}[A]-H_{1}-L_{\lambda}[A].
\end{equation}
We can again solve the modified constraint which, to leading order of $A$, becomes 
\begin{align}
c_{\mathbf{k}\alpha}= & \,\,u_{\alpha}^{*}(\mathbf{k})f_{\mathbf{k}}-\sum_{\mathbf{q}\mu}A_{\mathbf{q}}^{\mu}\mathcal{D}_{\mu}u_{\mathbf{k}+\frac{\mathbf{q}}{2}\alpha}^{*}f_{\mathbf{k+\mathbf{q}}}\nonumber \\
 & -\frac{1}{2}\sum_{\mathbf{q}\mu\nu}A_{\mathbf{q}}^{\mu}A_{\mathbf{-q}}^{\nu}G_{\mu\nu}(\mathbf{k})u_{\mathbf{k}\alpha}^{*}f_{\mathbf{k}}.\label{eq:ckA}
\end{align}
Here $\mathcal{D}_{\mu}=\partial_{\mu}+i\mathcal{A}_{\mu}$ is the
covariant derivative with Berry connection $\mathcal{A}_{\mu}(\mathbf{k})=i\langle u_{\mathbf{k}}\vert\partial_{\mu}u_{\mathbf{k}}\rangle$,
and $G_{\mu\nu}(\mathbf{k})$ is the quantum metric, $G_{\mu\nu}(\mathbf{k}) =\mathrm{Re}\langle\partial_{\mu}u_{\mathbf{k}}\vert\left(1-\vert u_{\mathbf{k}}\rangle\langle u_{\mathbf{k}}\vert\right)\vert\partial_{\nu}u_{\mathbf{k}}\rangle$.
The covariant derivative $\mathcal{D}_{\mu}$ appears due to the gauge
redundancy $u_{\mathbf{k}\alpha}\rightarrow u_{\mathbf{k}\alpha}e^{i\phi(\mathbf{k})}$.
The solution in Eq.~(\ref{eq:ckA}) captures the response of a Bloch
wave to a gauge potential $A_{\mathbf{q}}(t)$. 
An effective theory can be derived by
integrating out the fields $\lambda_{\mathbf{k}\alpha}$ and $c_{\mathbf{k}\alpha}$,
with $Z[A]=\int Df_{\mathbf{k}}Df_{\mathbf{k}}^{\dagger}e^{i\int dtL[f,A]}$
where the Lagrangian $L[f,A]$ becomes
\begin{equation}
L[f,A]=\sum_{\mathbf{k}}f_{\mathbf{k}}^{\dagger}i\partial_{t}f_{\mathbf{k}}^{}-H_{0}[f,A]-H_{1}[f,A],\label{eq:LfA}
\end{equation}
with $H_{0}[f,A]=\sum_{\mathbf{k}}\epsilon_{\mathbf{k}}f_{\mathbf{k}}^{\dagger}f_{\mathbf{k}}^{}-\sum_{\mathbf{k}\mu}A_{\mathbf{q}}^{\mu}\partial_{\mu}\epsilon_{\mathbf{k}}f_{\mathbf{k}}^{\dagger}f_{\mathbf{k}+\mathbf{q}}^ {}+\frac{1}{2}\sum_{\mathbf{k}\mu\nu}A_{\mathbf{q}}^{\mu}A_{-\mathbf{q}}^{\nu}\partial_{\mu}\partial_{\nu}\epsilon_{\mathbf{k}}f_{\mathbf{k}}^{\dagger}f_{\mathbf{k}}^ {}$.
The interaction term $H_{1}[f,A]$ is obtained by substituting the solution in Eq.~\eqref{eq:ckA} into the original $H_{1}$. 
The effective theory in Eq.~\eqref{eq:LfA} encodes band dispersion and Bloch wave changes by the gauge potential for Wannier-obstructed bands. 
We will derive the current operator based on the effective theory.

\emph{Quantum geometric current from interactions.---}
We first examine currents from the free term $H_0$. From Eq.~\eqref{eq:LfA},
it is easy to find the paramagnetic current and diamagnetic current
tensor from $H_0$, respectively $j_{\mathrm p,\mathbf{q}}^{\mu}=\sum_{\mathbf{k}}\partial_{\mu}\epsilon_\mathbf{k}f_{\mathbf{k}}^{\dagger}f_{\mathbf{k}+\mathbf{q}}^{}$,
and $K_{\mu\nu}=\sum_{\mathbf{k}}\partial_{\mu}\partial_{\nu}\epsilon_\mathbf{k}f_{\mathbf{k}}^{\dagger}f_{\mathbf{k}}^{}$ with the projected density operator $\rho_{\mathbf q}=\sum_{\mathbf k}f_{\mathbf k}^\dagger f_{\mathbf k+\mathbf q}$. 
The interaction term $H_{1}[f,A]$ can also contribute to the paramagnetic current $j_{\mathrm{QG},\mathbf{q}}^{\mu}$
and diamagnetic current tensor $K_{\mathrm{QG},\mu\nu}$, 
\begin{equation}
\!\!\!\!\!\! j_{\mathrm{QG},\mathbf{q}}^{\mu}=\frac{\delta H_{1}[f,A]}{\delta A_{-\mathbf{q}}^{\mu}}\vert_{A\rightarrow0},~K_{\mathrm{QG},\mu\nu}=\frac{\delta^{2}H_{1}[f,A]}{\delta A_{\mathbf{q}}^{\mu}\delta A_{-\mathbf{q}}^{\nu}}\vert_{A\rightarrow0},\label{eq:QGcurrent}
\end{equation}
even though $H_1$ does not contribute to the current operator in the full Hilbert space. 
Obviously, both $j_{\mathrm{QG},\mathbf{q}}^{\mu}$ and $K_{\mathrm{QG},\mu\nu}$
in Eq.~(\ref{eq:QGcurrent}), which we call QG currents,
are proportional to the interaction strength. The QG current encodes the dynamics from the interactions in the subspace $\mathbb H^\mathrm{sub}$. This is similar to the lowest Landau level \cite{1993IJMPB...7.4389M,1994MPLB....8.1065R}.
For a specific system, such as a flat-band superconductor, one can simplify the QG current expression by employing the ground state ansatz, as illustrated in the examples below.

One may question whether the Peierls substitution can be applied to
a band with a topological obstruction, where long-range hopping terms exist in the tight-binding models. There are two key observations. First, the free Hamiltonian $H_{0}$ does not contribute to QG current, and the related current only depends on band dispersion. 
Second, the projector in real space $\mathcal P_{\alpha\beta}(\mathbf r,\mathbf r^\prime ) = \sum _{\mathbf k} e^{i\mathbf k\cdot (\mathbf r-\mathbf r^\prime)} \mathcal P_{\alpha\beta}(\mathbf k)$ decreases exponentially for a large distance $|\mathbf r-\mathbf r^\prime|$ in any dimensions \cite{1964PhRv..135..685C,1983CMaPh..91...81N,1998PhRvB..58.3501G,1999PhRvL..82.2127I,2001PhRvL..86.5341H}, leading to a short-range effective interaction \cite{NOTE}. One typical example is the lowest Landau level, where the projected Coulomb interaction gets suppressed by a Gaussian decay factor.

\emph{Example I: Diamagnetic response in a flat-band
SC.---}  We revisit the SC in a flat-band system which possesses finite quantum geometry  \cite{2015NatCo68944P,PhysRevB.95.024515,2024PhRvL.132b6002C} from the perspective of the electromagnetic response. 
We consider the interaction
term: $H_{1}=-\frac{U}{N}\sum_{\mathbf{k}\mathbf{k}^{\prime}\mathbf{p}}\Delta_{\mathbf{k}}^{\dagger}(\mathbf{p})\Delta_{\mathbf{k}^{\prime}}^{}(\mathbf{p})$
with the pairing operator $\Delta_{\mathbf{k}}(\mathbf{p})=\sum_{\alpha}c_{-\mathbf{k}\alpha\downarrow}c_{\mathbf{k}+\mathbf{p}\alpha\uparrow}$.
We  examine its gauge coupling 
\begin{align}
\Delta_{\mathbf{k}}(\mathbf{p})\rightarrow & \,\,\Delta_{\mathbf{k}}(\mathbf{p})+\sum_{\mathbf{q}\mu}A_{\mathbf{q}}^{\mu}\mathcal{D}_{\mathbf{q}}^{\mu}\left(\gamma_{\mathbf{k}}^{*}(\mathbf{p})\right)\overline{\Delta}_{\mathbf{k}}(\mathbf{q}+\mathbf{p})\nonumber \\
 & -\sum_{\mathbf{q}\mu\nu}A_{\mathbf{q}}^{\mu}A_{-\mathbf{q}}^{\nu}G_{\mu\nu}(\mathbf{k})\overline{\Delta}_{\mathbf{k}}(\mathbf{p}),
\end{align}
with $\overline{\Delta}_{\mathbf{k}}(\mathbf{p})\equiv f_{\mathbf{-k}\downarrow}f_{\mathbf{k}+\mathbf{p}\uparrow}$
and $\gamma_{\mathbf{k}}(\mathbf{p})\equiv\langle u_{\mathbf{k}}\vert u_{\mathbf{k}+\mathbf{p}}\rangle$.
For convenience, we abbreviate $\mathcal{D}_{\mathbf{q}}^{\mu}\left(\langle u_{\mathbf{k}}\vert u_{\mathbf{k^{\prime}}}\rangle\right)\equiv\langle\mathcal{D}_{\mu}u_{\mathbf{k}+\frac{\mathbf{q}}{2}}\vert u_{\mathbf{k}^{\prime}}\rangle+\langle u_{\mathbf{k}}\vert\mathcal{D}_{\mu}u_{\mathbf{k}^{\prime}+\frac{\mathbf{q}}{2}}\rangle$. 
Then the paramagnetic current and diamagnetic current tensor are both proportional to the interaction,
\begin{small}
\begin{align}
\!\!\!j_{\mathrm{p},\mathbf{q}}^{\mu} & =\!-\frac{U}{N}\sum_{\mathbf{kk}^{\prime}\mathbf{p}}\gamma_{\mathbf{k}^{\prime}}(\mathbf{p})\mathcal{D}_{\mathbf{q}}^{\mu}\left(\gamma_{\mathbf{k}}^{*}(\mathbf{p})\right)\overline{\Delta}_{\mathbf{k}^{\prime}}^{\dagger}(\mathbf{p})\overline{\Delta}_{\mathbf{k}}(\mathbf{p}+\mathbf{q})+(\mathbf{k}\leftrightarrow\mathbf{k}^{\prime}),\\
K_{\mu\nu} & =\! \frac{U}{N}\sum_{\mathbf{k}\mathbf{k}^{\prime}\mathbf{p}}\left[G_{\mu\nu}(\mathbf{k})\gamma_{\mathbf{k}^{\prime}}(\mathbf{p})+G_{\mu\nu}(\mathbf{k}^{\prime})\gamma_{\mathbf{k}}^{*}(\mathbf{p})\right]\overline{\Delta}_{\mathbf{k}^{\prime}}^{\dagger}(\mathbf{p})\overline{\Delta}_{\mathbf{k}}(\mathbf{p}).
\end{align}
\end{small}
The Meissner effect can be inferred from the total current in linear response,
$\langle j_{\mathbf{q}}^{\mu}\rangle=-\sum_{\nu}D_s^{\mu\nu}A_{\mathbf{q}}^{\nu}$ 
where the superfluid weight is given by $D_s^{\mu\nu}= R_{\mu\nu}(\mathbf{q},t) + \langle K_{\mu\nu}\rangle $ with the paramagnetic current response tensor $R_{\mu\nu}(\mathbf{q},t)=-\langle[j_{\mathrm{p},\mathbf{q}}^{\mu}(t),j_{\mathrm{p},-\mathbf{q}}^{\nu}(0)]\rangle$.
The response can be evaluated based on the mean-field order parameter $\Delta=1/N\sum_{\mathbf{k}}\langle f_{-\mathbf{k}\downarrow}f_{\mathbf{k}\uparrow}\rangle$, which determines a BCS ground state \cite{SM}. At leading order in $\Delta$, we can decompose the paramagnetic current,
$j_{\mathrm{p},\mathbf{q}}^{\mu}=2U\Delta\sum_{\mathbf{k}}\mathcal{D}_{\mathbf{q}}^{\mu}\left(\langle u_{\mathbf{k}}\vert u_{\mathbf{k}}\rangle\right)[f_{-\mathbf{k}\downarrow}f_{\mathbf{k}+\mathbf{q}\uparrow}-f_{\mathbf{k}\uparrow}^{\dagger}f_{-\mathbf{k}-\mathbf{q}\downarrow}^{\dagger}]$, 
which vanishes when $q\rightarrow0$. 
In SM, we show that $R_{\mu\nu}(\mathbf{q},t)$
vanishes even at finite temperatures where the quasiparticle state
populations are nonzero. Thus, the current response at finite temperature is only determined by the diamagnetic response $\langle K_{\mu\nu}\rangle=2U\Delta^{2}\bar{G}_{\mu\nu}$
with $\bar{G}_{\mu\nu}$ being the averaged quantum metric over the first Brillouin
zone. This is consistent with Refs.~\cite{2015NatCo68944P,PhysRevB.95.024515,2024PhRvL.132b6002C} from different approaches. Therefore, the decrease of superfluid weight can be attributed to the shrinking of the pairing gap at finite temperature \cite{SM}. 
More importantly, the absence of paramagnetic current response at finite temperature indicates that the quasiparticle excitations do not carry dissipative currents, which can provide a mechanism behind the suppression of non-equilibrium quasiparticle transport observed in Ref.~\cite{2023PhRvL.130u6003P}. 
By contrast, in the two-fluid model of a conventional superconductor, dissipation is always present at finite temperatures due to the quasiparticle current~\cite{tinkham1996introduction}.

Concerns may arise regarding the validity of linear response theory in a flat-band SC state, where only one energy scale exists. We note that the amplitude of the form factor $\gamma_{\mathbf{k}}^{*}(\mathbf{p})\gamma_{\mathbf{k}^{\prime}}^ {}(\mathbf{p})$ which appears in the projected interaction term $H_{I}\rightarrow-\frac{U}{N}\sum_{\mathbf{k}\mathbf{k}^{\prime}\mathbf{p}}\gamma_{\mathbf{k}}^{*}(\mathbf{p})\gamma_{\mathbf{k}^{\prime}}(\mathbf{p})\overline{\Delta}_{\mathbf{k}}^{\dagger}(\mathbf{p})\overline{\Delta}_{\mathbf{k}^{\prime}}(\mathbf{p})$ is smaller than $1$. 
Therefore, it is likely still reasonable to conduct the linear response analysis.

\emph{Example II: Drude weight in a flat-band metal.---}
As the second example, we investigate the delocalization effect in
a flat-band metal. Ref.~\cite{2024arXiv240709599A} showed (through a diagrammatic expansion in the full Hilbert space) that there is a finite Drude weight in a flat-band metal with a non-zero quantum metric when the temperature is much larger than the interaction strength.
We  consider an 
interaction $H_{I}=\frac{1}{N}\sum_{\mathbf{k\mathbf{k}^{\prime}p}}V_{\mathbf{p}}\rho_{\mathbf{k^{\prime}}}^{\dagger}(\mathbf{p})\rho_{\mathbf{k}}(\mathbf{p})$
with $\rho_{\mathbf{k}}^ {}(\mathbf{p})=\sum_{\alpha}c_{\mathbf{k}\alpha}^{\dagger}c_{\mathbf{k}+\mathbf{p}\alpha}^{}$. 
With the generalized Peierls substitution, we  have the paramagnetic
current in response to a gauge potential $A_{t}^{\mu}$ of an electric field, 
\begin{align}
j_{\mathrm{p},t}^{\mu} & =\frac{1}{N}\sum_{\mathbf{k}\mathbf{k}^{\prime}\mathbf{p}}V_{\mathbf{p}}\mathcal{J}_{\mathbf{k}\mathbf{k}^{\prime}}(\mathbf{p})\bar{\rho}_{\mathbf{k}^{\prime}}^{\dagger}(\mathbf{p})\bar{\rho}_{\mathbf{k}}(\mathbf{p}),
\label{eq:Jcurrent}
\end{align}
with $\mathcal{J}_{\mathbf{k}\mathbf{k}^{\prime}}^{\mu}(\mathbf{p})  =\mathcal{D}_{\mathbf{0}}^{\mu}(\gamma_{\mathbf{k}^{\prime}}^{*}(\mathbf{p}))\gamma_{\mathbf{k}}(\mathbf{p})+\gamma_{\mathbf{k}^{\prime}}^{*}(\mathbf{p})\mathcal{D}_{\mathbf{0}}^{\mu}(\gamma_{\mathbf{k}}(\mathbf{p}))$.
When the temperature is much larger than the
bandwidth and the interaction strength, we can ignore the finite band dispersion from the Hartree-Fock decomposition of the interactions. In  linear response \cite{SM}, we can evaluate the Drude weight
\begin{align}
D_{\mu\nu}(T)= & \frac{\pi(\bar\nu(1-\bar\nu))^{2}}{T}\left[\int_{\mathbf{k\mathbf{k^{\prime}p}}}\vert V_{\mathbf{p}}\vert^{2}\mathcal{J}_{\mathbf{k}\mathbf{k}^{\prime}}^{\mu}(\mathbf{p})\mathcal{J}_{\mathbf{k}\mathbf{k}^{\prime}}^{\nu}(-\mathbf{p})\right.\nonumber \\
 & \left.-\int_{\mathbf{k}\mathbf{p}\mathbf{p}^{\prime}}V_{\mathbf{p}}V_{-\mathbf{p}^{\prime}}\mathcal{J}_{\mathbf{k}\mathbf{k}}^{\mu}(\mathbf{p})\mathcal{J}_{\mathbf{k}\mathbf{k}}^{\nu}(-\mathbf{p}^{\prime})\right],
\end{align} 
with $\bar\nu$ being the filling factor. In the special case of the potential $V_{\mathbf{p}}=V\delta_{\mathbf p,\mathbf 0}$, the Drude weight is related to
the Berry curvature variance, which is consistent with Ref.~\cite{2024arXiv240709599A}. Physically, the current $j_{\mathrm{p},t}^{\mu}$ describes a cooperative hopping process of two particles, which does not have a one-body operator correspondence. At low temperatures, the dispersion induced by the interaction becomes relevant, a problem we leave for future studies. 
Moreover, it will be interesting to consider the conductivity in the twisted bilayer graphene system where the narrow band possesses a fintie quantum geomery and the QG current can be induced by interactions as in Eq.~\eqref{eq:Jcurrent}.

\emph{Example III: Delocalisation from disorder.---} 
We further show that QG currents from disorder can induce a tendency towards delocalization in a flat-band system, in the absence of any conductivity in the clean limit \cite{2023PhRvB.108o5108H}. 
For this purpose, we can consider an on-site disorder in two spatial dimensions with $H_{1}=\sum_{\mathbf{r}\alpha}w_{\mathbf{r}}c_{\mathbf{r}\alpha}^{\dagger}c_{\mathbf{r}\alpha}$,
where the disorder satisfies $\overline{w_{\mathbf{r}}w_{\mathbf{r}^{\prime}}}=\gamma^{2}\delta_{\mathbf{r}\mathbf{r}^{\prime}}$ 
with $\overline{\cdot}$ denoting the disorder averaging.
Our generalized Peierls substitution to couple to $A_{t}^\mu$ of an electric field yields
\begin{equation}
\!\!\! L=\sum_{\mathbf{k}}if_{\mathbf{k}}^{\dagger}\partial_{t}f_{\mathbf{k}}^{}-\sum_{\mathbf{kp}}w_{\mathbf{p}}\langle u_{\mathbf{k}+\mathbf{p}}\vert u_{\mathbf{k}}\rangle f_{\mathbf{k}}^{\dagger}f_{\mathbf{k}+\mathbf{p}}^{}-A_{t}^{\mu}j_{\mathrm{p,t}}^{\mu},
\end{equation}
with $w_{\mathbf{p}}=1/N\sum_{\mathbf{r}}w_{\mathbf{r}}e^{i\mathbf{p}\cdot\mathbf{r}}$. Here the paramagnetic current 
\begin{equation}
j_{\mathrm{p},t}^{\mu}=\sum_{\mathbf{kp}}w_{\mathbf{p}}\mathcal{J}_{\mathbf{k}}^{\mu}(\mathbf{p})f_{\mathbf{k}}^{\dagger}f_{\mathbf{k}+\mathbf{p}}^{},
\label{eq:Jdis}
\end{equation}
with $\mathcal{J}_{\mathbf{k}}^{\mu}(\mathbf{p})=\langle u_{\mathbf{k+p}}\vert\mathcal{D}_{\mu}u_{\mathbf{k}}\rangle+\langle\mathcal{D}_{\mu}u_{\mathbf{k+p}}|u_{\mathbf{k}}\rangle$. This paramagnetic current $j_{\mathrm p,t}^\mu$ depends on the disorder configuration. 

In a flat-band system, introducing disorder can destroy the destructive quantum interference underpinning the vanishing group velocity. We note two effects of disorder in a flat-band system. First, disorder can lead to a finite bandwidth in the disorder-averaged Green function $\overline{G_{\mathbf{k}}}=\frac{1}{\omega/2-i\gamma}$ for $\vert\omega\vert\ll\gamma$ from the Dyson equation \cite{SM}. Second, the disorder drives the QG current, Eq.~\eqref{eq:Jdis}, which is absent in the clean limit. 
Ignoring the quantum fluctuations from Diffuson and Cooperon, we estimate the finite static conductivity when the disorder is much weaker than the band gap,
$\sigma_{\mu\nu}\sim\int\frac{d^{2}\mathbf{k}d^{2}\mathbf{p}}{(2\pi)^{4}}\mathcal{J}_{\mathbf{k}}^{\mu}(\mathbf{p})\mathcal{J}_{\mathbf{k}}^{\nu*}(\mathbf{p}).$
Such delocalization effect can potentially be linked to inverse Anderson localization \cite{PhysRevLett.96.126401,2024arXiv241219056C}.
The above discussion describes a regime at scales 
below the Anderson localization length, which in two dimensions can be large for weak disorder \cite{PhysRevLett.42.673,2023PhRvB.107f4205S}, as well as  energy dependent when band topology protects a delocalised state. 

The resulting detailed phenomenology, possibly including a `finite-size metal-insulator transition', is an interesting object which we leave to future study. Experimentally, the Lieb lattice, already realised in cold atomic systems 15 years ago \cite{PhysRevB.81.041410,2015SciA1E0854T,2016OptL41.1435X,2014NJPh16f3061G,2015PhRvL.114x5504M,2015PhRvL.114x5503V,2016PhRvL.116r3902D,2017NatPh..13..672S}, could provide a platform for experimental investigations.

\emph{Concluding remarks.---} 
It is essential to consider changes in Bloch waves when studying the response of bands of Wannier obstructions to general perturbations, which are not limited to electromagnetic fields. 
We note that our projection method can be extended to cases without translational symmetry where one can enforce 
projection onto the descendants of the flat band through $L_{\lambda}=\int_{\mathbf{r}}\lambda_{\mathbf{r}}^{\dagger}(c_{\mathbf{r}}-\sum_{n}\psi_{n}^{*}(\mathbf{r})f_{n})+h.c.$, 
with $n$ labeling the eigenstate of the bands and 
$\psi_n(\mathbf r)$  the corresponding wave functions. The projection restricts electrons to be spatially correlated, such that $c_\mathbf r=\int_{\mathbf r^\prime} G(\mathbf r,\mathbf r^\prime) c_{\mathbf r^\prime} $ with $G(\mathbf r,\mathbf r^\prime)=\sum_n \psi_n^*(\mathbf r)\psi_n(\mathbf r^\prime)$.

Finally, based on the examples above, one may view the interaction-induced
QG currents as integral to Fermi liquid behavior \cite{PhysRevLett.93.206602,10.21468/SciPostPhys.19.1.014}.
This immediately raises the question of how to extend Landau's Fermi liquid theory to incorporate QG currents.

\acknowledgments{S. Chen is indebted to the discussion with
Patrick A. Lee, Wen Huang, Siddhartha Sarkar and Johannes Mitscherling. This work was supported by the Deutsche Forschungsgemeinschaft via Research Unit FOR 5522 (project-id 499180199), as well as the cluster of excellence
ct.qmat (EXC-2147, project-id 390858490).}

\bibliography{ref}

\begin{thebibliography}{87}
\bibitem [{NOT({\natexlab{a}})}]{NOTEMB}%
  \BibitemOpen
  \href@noop {} {\bibfield  {journal} {\bibinfo  {journal} {Given $m$ bands at
  the Fermi energy, the projection becomes $c_{\mathbf k\alpha}\rightarrow
  \sum_{m}u^*_{m\mathbf k\alpha} f_{m\mathbf k}$ with $\sum_\beta
  h_{\alpha\beta}(\mathbf k)u_{m\mathbf k\beta}= \epsilon_{m\mathbf
  k}u_{m\mathbf k\alpha}$. Similarly, we have $\mathcal M_{\alpha\beta}(\mathbf
  k)=\sum_m u^*_{m\mathbf k\alpha}u_{m\mathbf k\beta}$ for the kernel in
  anticommutors of $c_{\mathbf k \alpha}$-fermions and $L_{\lambda} =
  \sum_{\mathbf{k}\alpha}\lambda_{\mathbf{k}\alpha}^{\dagger}(c_{\mathbf{k}\alpha}-\sum_m
  u_{m\mathbf{k}\alpha}^{*}f_{m\mathbf{k}})+h.c.$}\ }
  ({\natexlab{a}})}\BibitemShut {NoStop}%
\bibitem [{NOT({\natexlab{b}})}]{NOTE}%
  \BibitemOpen
  \href@noop {} {\bibfield  {journal} {\bibinfo  {journal} {For example, given
  a Coulomb interaction $ H_I = \int d\mathbf{r} d\mathbf{r}' \rho_{\mathbf{r}}
  V(\mathbf{r} - \mathbf{r}') \rho_{\mathbf{r}'}$, the projected interaction
  $\tilde{V}_{\mathbf{r}, \mathbf{r}'}=V(\mathbf{r} - \mathbf{r}')C(\mathbf
  r,\mathbf{r}')$ can be interpreted in terms of the two-particle correlation
  functions $C(\mathbf r,\mathbf{r}')= \langle \Omega | \rho_{\mathbf{r}}
  \rho_{\mathbf{r}'} | \Omega \rangle $, where $ |\Omega\rangle $ represents
  the ground state of an insulator with fully filled bands. In the
  independent-electron approximation, the $C(\mathbf r,\mathbf{r}')$ exhibits
  exponential decay as the distance $|\mathbf{r} - \mathbf{r}'|$ increases,
  potentially suppressing even a bare long-range interaction}\ }
  ({\natexlab{b}})}\BibitemShut {NoStop}%
\end{thebibliography}

\onecolumngrid
\appendix

\section{Review on the Peierls substitution}
For the sake of completeness, we review the procedure of the Peierls
substitution. The Peierls substitution provides a systematical way
to study the response of the current in the presence of the electromagnetic field. To begin with, we consider a multi-band Hamiltonian which is
modeled by 
\begin{equation}
H_{0}=\sum_{\mathbf{r}_{ij}}\sum_{\alpha\beta}h_{\alpha\beta}(\mathbf{r}_{ij})c_{i\alpha}^{\dagger}c_{j\beta},\label{smeq:h0}
\end{equation}
where the summation over the distance $\mathbf{r}_{ij}$ is conducted
over the finite hopping integral $h_{\alpha\beta}(\mathbf{r}_{ij})$.
The indices $\alpha,\beta$ represent the orbital degree of freedoms.
The $c_{i\alpha}$($c_{i\alpha}^{\dagger}$) is the annihilation (creation)
operator on the site $\mathbf{r}_{i}$ and orbital $\alpha$. We consider
a slowly varying electromagnetic field $\mathbf{A}(\mathbf{r})$.
The Peierls substitution can introduce the coupling to the gauge potential
$\mathbf{A}(\mathbf{r})$ via the procedure, 
\begin{align}
 & h_{\alpha\beta}(\mathbf{r}_{ij})c_{i\alpha}^{\dagger}c_{j\beta}\exp\left[-i\int_{\mathbf{r}_{i}}^{\mathbf{r}_{j}}\mathbf{A}(\mathbf{r})\cdot d\mathbf{r}\right],\label{eq:phase}
\end{align}
As the gauge field $\mathbf{A}(\mathbf{r})$ is smooth over the hopping
distance $\mathbf{r}_{ij}$, which are some cases we are interested
in, such as the static conductivity, we can further expand the exponential
phase factor to the second order, 
\begin{align}
 & h_{\alpha\beta}(\mathbf{r}_{ij})c_{i\alpha}^{\dagger}c_{j\beta}\exp\left[-ie\int_{\mathbf{r}_{i}}^{\mathbf{r}_{j}}\mathbf{A}(\mathbf{r})\cdot d\mathbf{r}\right]\nonumber \\
\simeq & h_{\alpha\beta}(\mathbf{r}_{ij})c_{\alpha r_{j}}^{\dagger}c_{\beta r_{i}}-i\mathbf{r}_{ji}\cdot\mathbf{A}(\mathbf{R}_{ij})h_{\alpha\beta}(\mathbf{r}_{ij})c_{i\alpha}^{\dagger}c_{j\beta} -\frac{1}{2}\sum_{\mu\nu}\mathbf{r}_{ji,\mu}\mathbf{r}_{ji,\nu}A_{\mu}(\mathbf{R}_{ij})A_{\nu}(\mathbf{R}_{ij})h_{\alpha\beta}(\mathbf{r}_{ij})c_{i\alpha}^{\dagger}c_{j\beta},
\end{align}
with $\mathbf{R}_{ij}=\frac{\mathbf{r}_{j}+\mathbf{r}_{i}}{2}$. The
validity of the expansion relies on the short-ranged hopping integral.
Thus, one can drive the local current operator by taking the derivative
of the Hamiltonian $j_{\mu}(\mathbf{R}_{ij})=-\frac{\delta H}{\delta A_{\mu}(\mathbf{R}_{ij})}$,
which can be decomposed into the diamagnetic and paramagnetic currents. In the momentum space, we have the paramagnetic current
$j_{\mathrm p,\mathbf{q}}^{\mu}$ and the diamagnetic current tensor $K_{\mu\nu}$,
\begin{align}
j_{\mathrm p,\mathbf{q}}^{\mu} & =-\sum_{\alpha\beta}\sum_{\mathbf{k}}\partial_{\mu}h_{\alpha\beta}(\mathbf{k})c_{\alpha\mathbf{\mathbf{k}-\frac{\mathbf{q}}{2}}}^{\dagger}c_{\beta\mathbf{k}+\frac{\mathbf{q}}{2}},\\
K_{\mu\nu} & =\sum_{\alpha\beta}\sum_{\nu}\sum_{\mathbf{k}}\partial_{\mu}\partial_{\nu}h_{\alpha\beta}(\mathbf{k})c_{\alpha\mathbf{k}}^{\dagger}c_{\beta\mathbf{k}}^{},
\end{align}
where the indices $\mu$, $\nu$ denotes the spatial dimensions and
$h_{\alpha\beta}(\mathbf{k})=\sum_{\mathbf{r}_{ij}}h_{\alpha\beta}(\mathbf{r}_{ij})e^{i\mathbf{k}\cdot\mathbf{r}_{ij}}$. For
a multi-band system, the currents are composed of both intra-band
and inter-band components. 

Let's consider a much simpler case with only one conduction band around
the Fermi energy. This conduction band gets separated from other bands
with large band gaps $\Delta_{\mathrm{gap}}$. Such systems have 
attracted a large amount of focus and a prototypical example is the
Moir\'e material which features narrow bands at magic angles. In the
leading order, one may restrict the multi-orbital system into an effective
one band with a sub-Hilbert space via the projection operator $\mathcal{P}=\otimes_{\mathbf{p}}\vert u(\mathbf{p})\rangle\langle u(\mathbf{p})\vert$.
Under the band basis, we have 
\begin{align}
[j_{\mathrm p,\mathbf{q}}^{\mu}]_{nm}= & \sum_{\mathbf{k}}(\epsilon_{n}(\mathbf{k})-\epsilon_{m}(\mathbf{k}+\mathbf{q}))\langle\partial_{\mu}u_{n\mathbf{k}}\vert u_{m\mathbf{k}+\mathbf{q}}\rangle  +\sum_{\mathbf{k}}\partial_{\mu}\epsilon_{n}(\mathbf{k})\langle u_{n\mathbf{k}}\vert u_{n\mathbf{k}+\mathbf{q}}\rangle\delta_{n,m},
\end{align}
with $H_0|u_{n\mathbf k}\rangle = \epsilon_n(\mathbf k)|u_{n\mathbf k}\rangle $.
We point out that the inter-band components are proportional to the
energy difference. However, it does not indicate the higher bands have larger weights. The current operator is not invariant
under the gauge choice of the Wannier functions, that is, the phase
shift 
\begin{equation}
\vert u_{n\mathbf{k}}\rangle\rightarrow e^{i\phi_{n}(\mathbf{k})}\vert u_{n\mathbf{k}}\rangle.
\label{smeq:wannergauge}
\end{equation}
Of course, the physical quantities, like the density-density correlation function
should keep invariant under the gauge symmetry in Eq.~\eqref{smeq:wannergauge}. 

\section{Euler-Lagrange equation}
\subsection{Property of $\mathcal P_{\alpha\beta}(\mathbf k)$}
After projection, the fermions satisfy an unconventional anticommutator where $\mathcal P_{\alpha\beta}(\mathbf k)$ replaces the $\delta_{\alpha\beta}$. The kernel $\mathcal P_{\alpha\beta}$ is an analog of the delta function. First, it retains the reproducing property, 
\begin{equation}
    \sum_{\beta}\mathcal P_{\alpha\beta}(\mathbf k)\mathcal P_{\beta\gamma} (\mathbf k)= \mathcal P_{\alpha\gamma}(\mathbf k).
\end{equation}

One can further establish the kernel $\mathcal P_{\alpha \beta }(\mathbf r,\mathbf r^\prime)$ in the real space with 
\begin{equation}
\mathcal P_{\alpha \beta }(\mathbf r,\mathbf r^\prime)=\sum_\mathbf{k}e^{i\mathbf k\cdot (\mathbf r-\mathbf r^\prime)} u_\mathbf{k\alpha}^*u_{\mathbf k\beta}.
\end{equation}
which appears when we consider the anticommutator in real space  with the subspace $\mathbb H^\mathrm{sub}$
\begin{equation}
    \{c_{\mathbf r\alpha},c_{\mathbf r^\prime\beta}^\dagger\} = \mathcal P_{\alpha\beta}(\mathbf r,\mathbf r^\prime).
\end{equation}
The kernel $\mathcal P_{\alpha \beta }(\mathbf r,\mathbf r^\prime)$ describes the spatial correlation.

\subsection{Derivation on EL equation}
In the main text, we introduce the Lagrange multipliers $\lambda_{\mathbf{k}\alpha}$ and $\lambda_{\mathbf{k}\alpha}^{\dagger}$
to realize the band projection, which allows us to regard the fermion fields
$c_{\mathbf{k}\alpha}$ and $c_{\mathbf{k}\alpha}^{\dagger}$ to obey
the normal anti-commutativity relation. In this section, we give the
details on the Euler-Lagrange (EL) equation within
the subspace $\mathbb{H}^{\mathrm{sub}}$.

Given the action $S=\int dtL$ with 
\begin{equation}
L=\sum_{\mathbf{k}\alpha}c_{\mathbf{k}\alpha}^{\dagger}i\partial_{t}c_{\mathbf{k}\alpha}-H_{0}-H_{I}-L_{\lambda},
\end{equation}
the EL equation can be obtained from the functional derivative $\frac{\delta S}{\delta\phi}=0,\quad\phi\in\{c,c^{\dagger},\lambda,\lambda^{\dagger},f,f^{\dagger}\}$.
Then we have a list of equations,
\begin{align}
i\partial_{t}c_{\mathbf{k}\alpha} & =\sum_{\beta}h_{\alpha\beta}(\mathbf{k})c_{\mathbf{k}\beta}+\frac{\delta H_{I}}{\delta c_{\mathbf{k}\alpha}^{\dagger}}+\lambda_{\mathbf{k}\alpha},\label{eq:dc1}\\
i\partial_{t}c_{\mathbf{k}\alpha}^{\dagger} & =-\sum_{\beta}h_{\beta\alpha}(\mathbf{k})c_{\mathbf{k}\beta}^{\dagger}+\frac{\delta H_{I}}{\delta c_{\mathbf{k}\alpha}}-\lambda_{\mathbf{k}\alpha}^{\dagger},\label{eq:dc2}\\
c_{\mathbf{k}\alpha} & =u_{\alpha}^{*}(\mathbf{k})f_{\mathbf{k}},\quad c_{\mathbf{k}\alpha}^{\dagger}=u_{\alpha}(\mathbf{k})f_{\mathbf{k}}^{\dagger},\label{eq:cons1}\\
0 & =\sum_{\alpha}\lambda_{\mathbf{k}\alpha}^{\dagger}u_{\mathbf{k}\alpha}^{*}=\sum_{\alpha}\lambda_{\mathbf{k}\alpha}u_{\mathbf{k}\alpha}.\label{eq:cons2}
\end{align}
To get rid of the Lagrange multipliers $\lambda_{\mathbf{k}\alpha}$
and $\lambda_{\mathbf{k}\alpha}^{\dagger}$, we can multiply a factor
$\mathcal{M}_{\gamma\alpha}(\mathbf{k})$ in the both side of Eq.~(\ref{eq:dc1}) and then sum over the index $\alpha$ which leads
to 
\begin{equation}
i\partial_{t}c_{\mathbf{k}\gamma}=\sum_{\alpha\beta}h_{\gamma\beta}(\mathbf{k})c_{\mathbf{k}\beta}+\sum_{\alpha}\mathcal{P}_{\gamma\alpha}(\mathbf{k})\frac{\delta L_{I}}{\delta c_{\mathbf{k}\alpha}^{\dagger}}, \label{eq:el_c_m}
\end{equation}
where we have used the conditions
\begin{align}
 & \sum_{\alpha}\mathcal{P}_{\gamma\alpha}(\mathbf{k})\left[\lambda_{\mathbf{k}\alpha}-\sum_{\beta}\mathcal{P}_{\alpha\beta}(\mathbf{k})\lambda_{\mathbf{k}\beta}\right]\nonumber \\
= & \sum_{\gamma}\left[\sum_{\alpha}\mathcal{P}_{\gamma\alpha}(\mathbf{k})\lambda_{\mathbf{k}\alpha}-\sum_{\alpha\beta}\mathcal{P}_{\gamma\alpha}(\mathbf{k})\mathcal{P}_{\alpha\beta}(\mathbf{k})\lambda_{\mathbf{k}\beta}\right]\nonumber \\
= & \sum_{\gamma}\left[\sum_{\alpha}\mathcal{P}_{\gamma\alpha}(\mathbf{k})\lambda_{\mathbf{k}\alpha}-\sum_{\alpha\beta}\mathcal{P}_{\gamma\beta}(\mathbf{k})\lambda_{\mathbf{k}\beta}\right]\nonumber \\
= & 0,
\end{align}
and 
\begin{equation}
\sum_{\alpha}\mathcal{P}_{\gamma\alpha}(\mathbf{k})h_{\alpha\beta}(\mathbf{k})=h_{\gamma\beta}(\mathbf{k}).
\end{equation}
Take Eq.~(\ref{eq:el_c_m}) back to Eq.~(\ref{eq:dc1}), and we have
\begin{align}
\sum_{\beta}\mathcal{P}_{\alpha\beta}(\mathbf{k})\frac{\delta H_{I}}{\delta c_{\beta}^{\dagger}(\mathbf{k})} & =\frac{\delta H_{I}}{\delta c_{\alpha}^{\dagger}(\mathbf{k})}+\lambda_{\alpha}(\mathbf{k}),
\end{align}
One can find the solution 
\begin{align}
\text{\ensuremath{\lambda}}_{\mathbf{k}\alpha} & =\sum_{\beta}\left(\mathcal{P}_{\alpha\beta}(\mathbf{k})-\delta_{\alpha\beta}\right)\frac{\delta H_{I}}{\delta c_{\mathbf{k}\alpha}^{\dagger}},\label{eq:eta_sol}\\
\lambda_{\mathbf{k}\alpha}^{\dagger} & =\sum_{\beta}\frac{\delta H_{I}}{\delta c_{\mathbf{k}\alpha}}\left(\mathcal{P}_{\alpha\beta}(\mathbf{k})-\delta_{\alpha\beta}\right).
\end{align}
The Lagrange multipliers depend on the interaction $H_{I}$, indicating
the necessity of the interaction for the quantum geometry effect to manifest. Therefore, we have the
final EL equations
\begin{align}
i\partial_{t}c_{\mathbf{k}\alpha} & =\epsilon_{\mathbf{k}}c_{\mathbf{k}\alpha}+\sum_{\beta}\mathcal{P}_{\alpha\beta}(\mathbf{k})\frac{\delta H_{I}}{\delta c_{\mathbf{k}\beta}^{\dagger}},\label{eq:EL1}\\
-i\partial_{t}c_{\mathbf{k}\alpha}^{\dagger} & =\epsilon_{\mathbf{k}}c_{\mathbf{k}\alpha}^{\dagger}-\sum_{\beta}\frac{\delta H_{I}}{\delta c_{\mathbf{k}\beta}}\mathcal{P}_{\beta\alpha}(\mathbf{k}).\label{eq:EL2}
\end{align}
The matrix$\mathcal{P}_{\alpha\beta}(\mathbf{k})$ appears before
the interaction terms.

We can then derive the continuity equation. Generally, one may not
expect a contribution from terms like the density-density interaction
to the paramagnetic current. As such, the factor $\mathcal{P}_{\alpha\beta}(\mathbf{k})$
can enable an interaction-driven current. We define the operator $\rho_{\mathbf{k}\alpha}(\mathbf{q})=c_{\mathbf{k}+\mathbf{q}\alpha}^{\dagger}c_{\mathbf{k}\alpha}$.
Using the EL equations, we can calculate its time derivation,
\begin{align}
i\partial_{t}\rho_{\mathbf{k}\alpha}(\mathbf{q})= & \sum_{\mathbf{k}\beta}\left[c_{\mathbf{k}+\mathbf{q}\alpha}^{\dagger}h_{\alpha\beta}(\mathbf{k})c_{\mathbf{k}\beta}-c_{\mathbf{k}+\mathbf{q}\beta}^{\dagger}h_{\beta\alpha}(\mathbf{k}+\mathbf{q})c_{\mathbf{k}\alpha}-c_{\mathbf{k}+\mathbf{q}\alpha}^{\dagger}\mathcal{P}_{\alpha\beta}(\mathbf{k})\lambda_{\mathbf{k}\beta}-\lambda_{\mathbf{k}+\mathbf{q}\beta}^{\dagger}\mathcal{P}_{\beta\alpha}(\mathbf{k}+\mathbf{q})c_{\mathbf{k}\alpha}\right]. \label{eq:dtrho}
\end{align}
The first line in Eq.~(\ref{eq:dtrho}) leads to the current form
the targeted band dispersion, 
\begin{align}
 & \sum_{\alpha}\sum_{\beta}\left[c_{\mathbf{k}+\mathbf{q}\alpha}^{\dagger}h_{\alpha\beta}(\mathbf{k})c_{\mathbf{k}\beta}-c_{\mathbf{k}+\mathbf{q}\beta}^{\dagger}h_{\beta\alpha}(\mathbf{k}+\mathbf{q})c_{\mathbf{k}\alpha}\right] 
\rightarrow (\epsilon_{\mathbf{k}}-\epsilon_{\mathbf{k}+\mathbf{q}})f_{\mathbf{k}+\mathbf{q}}^{\dagger}f_{\mathbf{k}}.
\end{align}
The dynamics of the density in Eq.~(\ref{eq:dtrho}) is not explicitly
dependent on the interaction term $H_{I}$ as the density-density
interaction commutes with the density operator $[H_{I},c_{\mathbf{r}\alpha}^{\dagger}c_{\mathbf{r}\alpha}]=0$
in the full Hilbert space $\mathbb{H}$. Hence, we have a conservation
law within the subspace $\mathbb{H}^{\mathrm{sub}}$,
\begin{align}
-i\partial_{t}\rho(\mathbf{q})= & \sum_{\mathbf{k}}(\epsilon_{\mathbf{k}}-\epsilon_{\mathbf{k}+\mathbf{q}})f_{\mathbf{k}+\mathbf{q}}^{\dagger}f_{\mathbf{k}} +\sum_{\mathbf{k}}\sum_{\alpha\beta}\left[c_{\alpha}^{\dagger}(\mathbf{k}+\mathbf{q})\mathcal{P}_{\alpha\beta}(\mathbf{k})\frac{\delta H_{I}}{\delta c_{\beta}^{\dagger}(\mathbf{k})}+\frac{\delta H_{I}}{\delta c_{\alpha}(\mathbf{k}+\mathbf{q})}\mathcal{P}_{\alpha\beta}(\mathbf{k}+\mathbf{q})c_{\beta}(\mathbf{k})\right],
\end{align}
with $\rho(\mathbf{q})=\sum_{\mathbf{k}\alpha}\rho_{\mathbf{k}\alpha}(\mathbf{q})$.
The current operator can be reached by expanding the right-hand to
the form $\mathbf{q}\cdot\mathbf{j}_{\mathbf{q}}$.



\section{Example: flat-band superconductor}

We will give the details of the current operator in a flat-band superconductor.
For convenience, we consider the attractive interaction 
\begin{equation}
H_{I}=-\frac{U}{N}\sum_{\mathbf{k}\mathbf{k}^{\prime}\mathbf{p}}\Delta_{\mathbf{k}}^{\dagger}(\mathbf{p})\Delta_{\mathbf{k}^{\prime}}(\mathbf{p})
\end{equation}
with the Cooper pair operator $\Delta_{\mathbf{k}}(\mathbf{p})=\sum_{\alpha}c_{\mathbf{-k}\alpha\downarrow}c_{\mathbf{k}+\mathbf{p}\alpha\uparrow}$.
We assume the time-reversal symmetry, with $u_{-\mathbf{k}\downarrow}^{*}=u_{\mathbf{k}\uparrow}=u_{\mathbf{k}}$.
To consider the finite temperature, we can construct the path integral
under the imaginary time 
\begin{equation}
Z=\int Dc^{\dagger}DcD\lambda^{\dagger}D\lambda Df^{\dagger}Df\exp\left(-\int_{0}^{\beta}d\tau L\right)
\end{equation}
with the Lagrangian,
\begin{equation}
L=\sum_{\mathbf{k}\alpha\sigma}f_{\mathbf{k}\alpha\sigma}^{\dagger}\partial_{\tau}f_{\mathbf{k}\alpha\sigma}+H_{0}+H_{I}+L_{\lambda},
\end{equation}
where 
\begin{align}
H_{0} & =\sum_{\alpha\beta}\sum_{\mathbf{k}\sigma}h_{\alpha\beta}(\mathbf{k})c_{\mathbf{k}\alpha\sigma}^{\dagger}c_{\mathbf{k}\alpha\sigma},\\
L_{\lambda} & =\sum_{\mathbf{k}}\lambda_{\mathbf{k}\alpha\sigma}^{\dagger}(c_{\mathbf{k}\alpha\sigma}-u_{\mathbf{k}\alpha\sigma}^{*}f_{\mathbf{k\sigma}})+h.c..
\end{align}
We first analyze the mean-field without external perturbation. Obviously,
we have the constraint after integrating out the Lagrange multipliers
$\lambda$ and $\lambda^{\dagger}$
\begin{align}
c_{\mathbf{k}\alpha\sigma} & =u_{\mathbf{k}\alpha\sigma}^{*}f_{\mathbf{k\sigma}},\\
c_{\mathbf{k}\alpha\sigma}^{\dagger} & =u_{\mathbf{k}\alpha\sigma}f_{\mathbf{k\sigma}}^{\dagger}.
\end{align}
which gives rise to 
\begin{equation}
L=\sum_{\mathbf{k}\sigma}f_{\mathbf{k}\sigma}^{\dagger}\partial_{\tau}f_{\mathbf{k}\sigma}+\epsilon_{\mathbf{k}}f_{\mathbf{k}\sigma}^{\dagger}f_{\mathbf{k}\sigma}-\frac{U}{N}\sum_{\mathbf{k}\mathbf{k}^{\prime}\mathbf{p}}\gamma_{\mathbf{k}}^{*}(\mathbf{p})\gamma_{\mathbf{k}^{\prime}}(\mathbf{p})f_{\mathbf{k+p}\uparrow}^{\dagger}f_{-\mathbf{k}\downarrow}^{\dagger}f_{-\mathbf{k}^{\prime}\downarrow}f_{\mathbf{k}^{\prime}+\mathbf{p}\uparrow},
\end{equation}
with $\gamma_{\mathbf{k}}(\mathbf{p})=$$\langle u_{\mathbf{k}}\vert u_{\mathbf{k}+\mathbf{p}}\rangle$.
Our purpose is to consider the interaction effect, and we can assume
a flat-band limit $\epsilon_{\mathbf{k}}=0$.The mean-field theory
has been exhaustedly studied by introducing the mean-field ansatz
$\Delta=\frac{1}{N}\sum_{\mathbf{k}}\langle f_{-\mathbf{k}\downarrow}f_{\mathbf{k}\uparrow}\rangle$
which defines the BCS ground state
\begin{equation}
\vert\mathrm{BCS}\rangle=\prod_{\mathbf{k}}\left[\mathcal{U}_{\mathbf{k}}+\mathcal{V}_{\mathbf{k}}f_{\mathbf{k}\uparrow}^{\dagger}f_{-\mathbf{k}\downarrow}^{\dagger}\right]\vert0\rangle,
\end{equation}
with the Bogoliubov transformation
\begin{align}
f_{\mathbf{k}\uparrow}^{\dagger} & =\mathcal{U}_{\mathbf{k}}\gamma_{\uparrow}^{\dagger}(\mathbf{k})+\mathcal{V}_{\mathbf{k}}\gamma_{\downarrow}(-\mathbf{k}),\\
f_{-\mathbf{k}\downarrow} & =-\mathcal{V}_{\mathbf{k}}\gamma_{\uparrow}^{\dagger}(\mathbf{k})+\mathcal{U}_{\mathbf{k}}\gamma_{\downarrow}(-\mathbf{k}),
\end{align}
where we have chosen the gauge as such both $\mathcal{U}_{\mathbf{k}}$
and $\mathcal{V}_{\mathbf{k}}$ are real. In particular, in the flat-band
limit, $\mathcal{U}_{\mathbf{k}}=\mathcal{V}_{\mathbf{k}}=\frac{1}{\sqrt{2}}$.

We continue to give more analysis on the magnetic field effect. After the minimal coupling, the path function is 
\begin{equation}
Z[A,\phi]=\int Df_{\mathbf{k}}Df_{\mathbf{k}}^{\dagger}\exp\left[-\int_{0}^{\beta}d\tau\left(L+\sum_{\mu}j_{p,\mathbf{q}}^{\mu}A_{-\mathbf{q},t}^{\mu}+\frac{1}{2}\sum_{\mu\nu}K_{\mu\nu}A_{\mathbf{q},t}^{\mu}A_{-\mathbf{q},t}^{\nu}\right)\right],
\end{equation}
with 
\begin{align}
L & =f_{\mathbf{k}\sigma}^{\dagger}\partial_{\tau}f_{\mathbf{k}\sigma}+\epsilon_{\mathbf{k}}f_{\mathbf{k}\sigma}^{\dagger}f_{\mathbf{k}\sigma}-\frac{U}{N}\sum_{\mathbf{k}\mathbf{k}^{\prime}\mathbf{p}}\gamma_{\mathbf{k}}^{*}(\mathbf{p})\gamma_{\mathbf{k}^{\prime}}(\mathbf{p})f_{\mathbf{k+p}\uparrow}^{\dagger}f_{-\mathbf{k}\downarrow}^{\dagger}f_{-\mathbf{k}^{\prime}\downarrow}f_{\mathbf{k}^{\prime}+\mathbf{p}\uparrow},\\
j_{\mathrm{p},\mathbf{q}}^{\mu} & =-\frac{U}{N}\sum_{\mathbf{kk}^{\prime}\mathbf{p}}\gamma_{\mathbf{k}^{\prime}}(\mathbf{p})\mathcal{D}_{\mathbf{q}}^{\mu}\left(\gamma_{\mathbf{k}}^{*}(\mathbf{p})\right)\overline{\Delta}_{\mathbf{k}^{\prime}}^{\dagger}(\mathbf{p})\overline{\Delta}_{\mathbf{k}}(\mathbf{p}+\mathbf{q})+(\mathbf{k}\leftrightarrow\mathbf{k}^{\prime}),\\
K_{\mu\nu} & =\frac{U}{N}\sum_{\mathbf{k}\mathbf{k}^{\prime}\mathbf{p}}\left[G_{\mu\nu}(\mathbf{k})\gamma_{\mathbf{k}^{\prime}}(\mathbf{p})+G_{\mu\nu}(\mathbf{k}^{\prime})\gamma_{\mathbf{k}}^{*}(\mathbf{p})\right]f_{\mathbf{k+p}\uparrow}^{\dagger}f_{-\mathbf{k}\downarrow}^{\dagger}f_{-\mathbf{k}^{\prime}\downarrow}f_{\mathbf{k}^{\prime}+\mathbf{p}\uparrow},
\end{align}
where we abbreviate 
\begin{equation}
\mathcal{D}_{\mathbf{q}}^{\mu}\left(\langle u_{\mathbf{k}}\vert u_{\mathbf{k^{\prime}}}\rangle\right)\equiv\langle\mathcal{D}_{\mu}u_{\mathbf{k}+\frac{\mathbf{q}}{2}}\vert u_{\mathbf{k}^{\prime}}\rangle+\langle u_{\mathbf{k}}\vert\mathcal{D}_{\mu}u_{\mathbf{k}^{\prime}+\frac{\mathbf{q}}{2}}\rangle.
\end{equation}

\subsection{Mean-field theory under the gauge potential}

We further consider the mean-field theory on $L[A]$. Under the mean-field ansatz,
\begin{align}
\Delta(A) & =\frac{1}{N}\sum_{\mathbf{k}}\langle f_{-\mathbf{k}\downarrow}f_{\mathbf{k}\uparrow}\rangle
\end{align}
we decompose both the interaction term $H_{I}$ and $A_{\mathbf{p}}^{\mu}A_{-\mathbf{p}}^{\nu}K_{\mu\nu}$,
which gives rise to $L_{mf}$ with 
\begin{equation}
L_\mathrm{mf}=L_\mathrm{mf}[f]+L_\mathrm{mf}[f,A]
\end{equation}
where 
\begin{align}
L_\mathrm{mf}[f] & =\sum_{\mathbf{k}\sigma}\left[f_{\mathbf{k}\sigma}^{\dagger}\partial_{\tau}f_{\mathbf{k}\sigma}+\epsilon_{\mathbf{k}}f_{\mathbf{k}\sigma}^{\dagger}f_{\mathbf{k}\sigma}+U\left(\vert\Delta(A)\vert^{2}-\Delta(A) f_{-\mathbf{k}\downarrow}f_{\mathbf{k}\uparrow}-\Delta (A)f_{\mathbf{k}\uparrow}^{\dagger}f_{-\mathbf{k}\downarrow}^{\dagger}\right)\right],\\
L_\mathrm{mf}[f,A] & =-U\sum_{\mathbf{k}\sigma}A_{\mathbf{p}}^{\mu}A_{-\mathbf{p}}^{\nu}\left(G_{\mu\nu}(\mathbf{k})+\bar{G}_{\mu\nu}\right)\left(\vert\Delta(A)\vert^{2}-\Delta(A) f_{-\mathbf{k}\downarrow}f_{\mathbf{k}\uparrow}-\Delta(A) f_{\mathbf{k}\uparrow}^{\dagger}f_{-\mathbf{k}\downarrow}^{\dagger}\right).
\end{align}
Here $\bar{G}_{\mu\nu}=1/N\sum_{\mathbf{k}}G_{\mu\nu}(\mathbf{k})$
over the first Bouillon zone. Alternatively, we can have the mean-field
Hamiltonian , 
\begin{equation}
H_\mathrm{mf}[A]=\sum_{\mathbf{k}\sigma}\epsilon_{\mathbf{k}}f_{\mathbf{k}\sigma}^{\dagger}f_{\mathbf{k}\sigma}^{}+\sum_{\mathbf{k}}\sum_{\mu\nu}UA_{\mathbf{p}}^{\mu}A_{-\mathbf{p}}^{\nu}\left(G_{\mu\nu}(\mathbf{k})+\bar{G}_{\mu\nu}\right)\left(\vert\Delta(A)\vert^{2}-\Delta(A) f_{-\mathbf{k}\downarrow}f_{\mathbf{k}\uparrow}-\Delta(A) f_{\mathbf{k}\uparrow}^{\dagger}f_{-\mathbf{k}\downarrow}^{\dagger}\right).\label{eq:Hmf_A}
\end{equation}
The $A^{2}$ term also contributes to the mean-field Hamiltonian.
For a conventional dispersive band system, the $A^{2}$ term, i.e.
diamagnetic current, has the form $\frac{1}{m}A^{2}$ which will not
modify the mean-field order parameter. We emphasize that the order
parameter depends on the gauge field $A$. 

\subsubsection{Self-consistent equation}

As we are interested in the effect of the interaction, we can consider
a flat band $\epsilon_{\mathbf{k}}=\epsilon$. We can diagonalize
the mean-field Hamiltonian $H_\mathrm{mf}[A]$ in Eq.~(\ref{eq:Hmf_A}) by
the Bogoliubov transformation. We can diagonalize the mean-field Hamiltonian by the Bogoliubov transformation, and the quasiparticle
has the dispersion
\begin{equation}
E_{\mathbf{k}}(A)=\sqrt{(\epsilon-\mu(A))^{2}+U^{2}\Delta^{2}(A)\left[1-\sum_{\mu\nu}A_{\mathbf{p}}^{\mu}A_{-\mathbf{p}}^{\nu}(G_{\mu\nu}(\mathbf{k})+\bar{G}_{\mu\nu})\right]^{2}},
\end{equation}
where we keep the explicit dependence on the gauge field in the mean-field order parameter $\Delta(A)$ and the chemical potential $\mu(A)$. We
determine the $\Delta(A)$ and $\mu(A)$, we can make the expansion
\begin{align}
\Delta(A) & =(1+\sum_{\mu\nu}x_{\mu\nu}A_{\mathbf{p}}^{\mu}A_{-\mathbf{p}}^{\nu})\Delta+\mathcal{O}(A^{4}),\\
\mu(A) & =(1+\sum_{\mu\nu}y_{\mu\nu}A_{\mathbf{p}}^{\mu}A_{-\mathbf{p}}^{\nu})\mu+\mathcal{O}(A^{4}),
\end{align}
with $x_{\mu\nu}$ and $y_{\mu\nu}$ to be determined. Here, the $\Delta$ and $\mu$ satisfy the self-consistent equation and number equation without an external magnetic field, respectively,
 \begin{align}
 \frac{1}{U} & =\frac{\tanh\frac{\beta E}{2}}{2E},\\
 \nu & = 1-\frac{\epsilon-\mu}{E}\tanh\frac{\beta E}{2}.
 \end{align}
with $E=\sqrt{(\epsilon-\mu)^{2}+U^{2}\Delta^{2}}$ is the quasiparticle dispersion without external magnetic field. On
the other hand, the gap equation and number equation for the $\Delta(A)$
and $\mu(A)$ are 
\begin{align}
\frac{1}{U} & =\frac{1}{N}\sum_{\mathbf{k}}\frac{\tanh\frac{\beta E_{\mathbf{k}}(A)}{2}}{2E_{\mathbf{k}}(A)}\left[1-\sum_{\mu\nu} A_{\mathbf{p}}^{\mu}A_{-\mathbf{p}}^{\nu}(G_{\mu\nu}(\mathbf{k})+\bar{G}_{\mu\nu})\right], \\
\nu & =\frac{1}{N}\sum_{\mathbf{k}}\left(1-\frac{\epsilon-\mu(A)}{E_{\mathbf{k}}(A)}\tanh\frac{\beta E_{\mathbf{k}}(A)}{2}\right).
\end{align}
Expand the two equations to the second order, and we have the equations
for the $\alpha$ and $\gamma$, 
\begin{align*}
 & \sinh(\beta E)\left[\Delta^{2}U^{2}(2\bar{G}_{\mu\nu}+x_{\mu\nu}-y_{\mu\nu})-\epsilon^{2}(2\bar{G}_{\mu\nu}+x_{\mu\nu})\right]-\beta\Delta^{2}U^{2}(2\bar{G}_{\mu\nu}+x_{\mu\nu}-y_{\mu\nu})+\beta E^{3}x_{\mu\nu}=0,\\
 & \Delta^{2}U^{2}\sinh(\beta E)(2\bar{G}_{\mu\nu}+x_{\mu\nu}-y_{\mu\nu})+\beta E\left[\Delta^{2}U^{2}(y_{\mu\nu}-2\bar{G}_{\mu\nu})+(\epsilon-\mu)^{2}x_{\mu\nu}\right]=0.
\end{align*}
It is easy to find the solution 
\begin{align}
x_{\mu\nu} & =\frac{2\beta E^{3}\bar{G}_{\mu\nu}}{U^{2}\Delta^{2}(\beta E-\sinh(\beta E))},\\
y_{\mu\nu} & =-2\bar{G}_{\mu\nu}\frac{\epsilon-\mu}{\mu}.
\end{align}
In particular, at very low temperatures, $x\rightarrow0$, we can ignore
the dependence of $\Delta(A)$ on $A$. Therefore, up to the leading
order of $A$, we have
\begin{align}
\mu(A) & =\mu(1-2\bar{G}_{\mu\nu}A_\mu A_\nu),\\
\Delta(A) & =\Delta.
\end{align}

\subsubsection{Superfluid weight}
We can evaluate the grand potential $F_{0}[A]$ 
\begin{align}
F_{0}[A] & =\sum_{\mathbf{k}\sigma}\epsilon_{\mathbf{k}}\langle f_{\mathbf k\sigma}^{\dagger}f_{\mathbf k\sigma}\rangle+\sum_{\mathbf{k}}U\left[1-A_{\mathbf{p}}^{\mu}A_{-\mathbf{p}}^{\nu}(G_{\mu\nu}(\mathbf{k})+\bar{G}_{\mu\nu})\right]\left[\vert\Delta(A)\vert^{2}-\Delta(A)\langle f_{-\mathbf{k}\downarrow}f_{\mathbf{k}\uparrow}\rangle-\Delta(A)\langle f_{\mathbf{k}\uparrow}^{\dagger}f_{-\mathbf{k}\downarrow}^{\dagger}\rangle\right] \notag \\
 & =F_{0}+\delta F[A],
\end{align}
where  $F_0$ is the free energy without external gauge field, and $\delta F[A]$ is the deviation induced by the gauge potential,
\begin{equation}
\delta F[A]=\frac{1}{2}(D_{s,n}^{\mu\nu}+D_{s,p}^{\mu\nu})A_{\mathbf{p}}^{\mu}A_{-\mathbf{p}}^{\nu},
\end{equation}
with the superfluid weight,
\begin{align}
D_{s,p}^{\mu\nu} & =\frac{1}{N}\sum_{\mathbf{k}}U^2G_{\mu\nu}(\mathbf{k})\Delta\left[\langle f_{-\mathbf{k}\downarrow}f_{\mathbf{k}\uparrow}\rangle_{0}+\langle f_{\mathbf{k}\uparrow}^{\dagger}f_{-\mathbf{k}\downarrow}^{\dagger}\rangle_{0}\right]\nonumber \\
 & =2(U\Delta)^{2}\frac{1}{N}\sum_{\mathbf{k}}G_{\mu\nu}(\mathbf{k})\frac{\tanh\frac{\beta E_{\mathbf{k}}}{2}}{E_{\mathbf{k}}},
\end{align}
and 
\begin{align}
D_{s,n}^{\mu\nu} & =\frac{1}{N}\sum_{\mathbf{k}}2\epsilon_{\mathbf{k}}\frac{\partial\langle f_{\mathbf{k}\sigma}^{\dagger}f_{\mathbf{k}\sigma}\rangle}{\partial\mu}\frac{\partial\mu}{\partial(A_{\mathbf{p}}^{\mu}A_{-\mathbf{p}}^{\nu})}\nonumber \\
 & =\frac{1}{N}\sum_{\mathbf{k}}-2\bar{G}_{\mu\nu}\epsilon_{\mathbf{k}}\frac{\partial n_\mathrm{FD}(E_{\mathbf{k}})}{\partial\mu}\nonumber \\
 & =2\bar{G}\frac{\beta (\epsilon-\mu)^2}{\sqrt{(\epsilon-\mu)^{2}+(U\Delta)^{2}}}n_\mathrm{FD}(E)\left[1-n_\mathrm{FD}(E)\right].
\end{align}
The superfluid weight $D_{s,p}$ recovers the previous result. The $D_{s,n}$
arises from the change in chemical potential.
At very low temperature, $n_\mathrm{FD}(E)\left[1-n_\mathrm{FD}(E)\right]$
is zero, and thus there is no extra contribution from the fluctuations
of the chemical potential. 

Due to the gauge symmetry, after integrating the fluctuations $\psi_{\mathbf k}(\mathbf q)$ which occurs around the mean-field configuration,
\begin{align}
   \Delta_{\mathbf{k}}(\mathbf{q}) & =\Delta\delta_{\mathbf{q},\mathbf{0}}+\psi_{\mathbf{k}}(\mathbf{q}) ,
\end{align}
we will obtain the effective action 
\begin{equation}
S[A]=\frac{\beta}{2}\sum_{\mathbf q}D_{s}A_{\mathbf q}^{\bot}A_{-\mathbf q}^{\bot},
\end{equation}
where $A_{q}^{\bot}$ is the transverse component of the gauge field
$A$
\begin{equation}
\mathbf{A}_{q}^{\bot}=\mathbf{A}_{q}-\frac{\mathbf{q}}{q^{2}}\mathbf{q}\cdot\mathbf{A}_{q}.
\end{equation}

\subsection{Superfluid weight from the current-current correlation function}

We can calculate the superfluid weight by inspecting the current correlator, which determines the total current $\langle j^\mu{\mathbf q}\rangle $,
\begin{equation}
\langle j^{\mu}(\mathbf{q})\rangle=-\sum_{\nu}\left(R_{\mu\nu}(\mathbf{q},t)+\langle K_{\mu\nu}\rangle\right)A_{\mathbf{q},t}^{\nu}
\end{equation}
with 
\begin{align}
R_{\mu\nu}(\mathbf{q},t) & =-\langle[j_{p,\mathbf{q}}^{\mu}(t),j_{p,-\mathbf{q}}^{\nu}(0)]\rangle.
\end{align}
Under the BCS mean-field ansatz, we will evaluate the paramagnetic
and diamagnetic responses, respectively. We first apply the mean-field
ansatz on the diamagnetic current tensor by keeping the leading order,
\begin{align}
\langle K_{\mu\nu}\rangle & =\frac{U}{N}\sum_{\mathbf{k}\mathbf{k}^{\prime}\mathbf{p}}[G_{\mu\nu}(\mathbf{k})\gamma_{\mathbf{k}^{\prime}}(\mathbf{p})+G_{\mu\nu}(\mathbf{k}^{\prime})\gamma_{\mathbf{k}}(\mathbf{p})]\langle f_{\mathbf{k+p}\uparrow}^{\dagger}f_{-\mathbf{k}\downarrow}^{\dagger}f_{-\mathbf{k}^{\prime}\downarrow}f_{\mathbf{k}^{\prime}+\mathbf{p}\uparrow}\rangle\nonumber \\
 & = 2U\Delta^{2}\bar{G}_{\mu\nu}.
\end{align}
For the paramagnetic current, based on the mean-field ansatz, we can further decompose the paramagnetic current 
\begin{align}
j_{\mathrm{p},\mathbf{q}}^{\mu}= ~ & 2U\sum_{\mathbf{k}}\mathcal{D}_{\mathbf{q}}^{\mu}(\langle u_{\mathbf{k}}\vert u_{\mathbf{k}}\rangle)\left(f_{-\mathbf{k}\downarrow}f_{\mathbf{k}+\mathbf{q}\uparrow}-f_{\mathbf{k}\uparrow}^{\dagger}f_{-\mathbf{k}-\mathbf{q}\downarrow}^{\dagger}\right)\nonumber \\
= ~& 4U\sum_{\mathbf{k}}\mathcal{D}_{\mathbf{q}}^{\mu}(\langle u_{\mathbf{k}}\vert u_{\mathbf{k}}\rangle)\left\{(\mathcal{V}_{\mathbf{k}}\mathcal{V}_{\mathbf{k+\mathbf{q}}}+\mathcal{U}_{\mathbf{k}}\mathcal{U}_{\mathbf{k}+\mathbf{q}})\left[\gamma_{\downarrow}^{\dagger}(-\mathbf{k}-\mathbf{q})\gamma_{\uparrow}^{\dagger}(\mathbf{k})-\gamma_{\uparrow}(\mathbf{k}+\mathbf{q})\gamma_{\downarrow}(-\mathbf{k})\right]\right.\nonumber \\
 ~& \left.+(\mathcal{U}_{\mathbf{k}}\mathcal{V}_{\mathbf{k+\mathbf{q}}}-\mathcal{U}_{\mathbf{k}+\mathbf{q}}\mathcal{V}_{\mathbf{k}})\left[\gamma_{\uparrow}^{\dagger}(\mathbf{k})\gamma(\mathbf{k}+\mathbf{q})+\gamma_{\downarrow}^{\dagger}(-\mathbf{k}-\mathbf{q})\gamma_{\downarrow}(-\mathbf{k})\right]\right\} \nonumber\\
= ~& 4U\sum_{\mathbf{k}}\mathcal{D}_{\mathbf{q}}^{\mu}(\langle u_{\mathbf{k}}\vert u_{\mathbf{k}}\rangle)\left\{p_{\mathbf{k},\mathbf{k}+\mathbf{q}}\left[\gamma_{\downarrow}^{\dagger}(-\mathbf{k}-\mathbf{q})\gamma_{\uparrow}^{\dagger}(\mathbf{k})-\gamma_{\uparrow}(\mathbf{k}+\mathbf{q})\gamma_{\downarrow}(-\mathbf{k})\right]\right.\nonumber \\
 ~& \left.+\ell_{\mathbf{k},\mathbf{k}+\mathbf{q}}\left[\gamma_{\uparrow}^{\dagger}(\mathbf{k})\gamma(\mathbf{k}+\mathbf{q})+\gamma_{\downarrow}^{\dagger}(-\mathbf{k}-\mathbf{q})\gamma_{\downarrow}(-\mathbf{k})\right]\right\},
\end{align}
 with $\mathcal{D}_{\mathbf{q}}^{\mu}(\langle u_{\mathbf{k}}\vert u_{\mathbf{k}}\rangle)=\langle\mathcal{D}_{\mu}u_{\mathbf{k}+\frac{\mathbf{q}}{2}}\vert u_{\mathbf{k}}\rangle+\langle u_{\mathbf{k}}\vert\mathcal{D}_{\mu}u_{\mathbf{k}+\frac{\mathbf{q}}{2}}\rangle$.
Here $p_{\mathbf{k},\mathbf{k}+\mathbf{q}}=\mathcal{V}_{\mathbf{k}}\mathcal{V}_{\mathbf{k+\mathbf{q}}}+\mathcal{U}_{\mathbf{k}}\mathcal{U}_{\mathbf{k}+\mathbf{q}}$
and $\ell_{\mathbf{k},\mathbf{k}+\mathbf{q}}=\mathcal{U}_{\mathbf{k}}\mathcal{V}_{\mathbf{k+\mathbf{q}}}-\mathcal{U}_{\mathbf{k}+\mathbf{q}}\mathcal{V}_{\mathbf{k}}$
are the coherence factors. 

At $T=0$, the BCS ground state has no quasiparticle excitations and we have 
\begin{align}
 & \langle[j_{p,q}^{\mu}(t),j_{p,-q}^{\nu}(0)]\rangle\nonumber \\
= & \sum_{\mathbf{k}}\mathcal{D}_{\mathbf{q}}^{\mu}(\langle u_{\mathbf{k}}\vert u_{\mathbf{k}}\rangle)\left[\mathcal{D}_{\mathbf{q}}^{\nu}(\langle u_{\mathbf{k}}\vert u_{\mathbf{k}}\rangle)\right]^{*}p_{\mathbf{k},\mathbf{k}+\mathbf{q}}^{2}\left(\frac{1}{\omega-(E_{\mathbf{k}}+E_{\mathbf{k}+\mathbf{q}})+i\eta}-\frac{1}{\omega+(E_{\mathbf{k}}+E_{\mathbf{k}+\mathbf{q}})+i\eta}\right).
\end{align}
In the DC limit, $\omega=0$ and we have 
\begin{equation}
R_{\mu\nu}(\mathbf{q},\omega=0)=\sum_{\mathbf{k}}\mathcal{D}_{\mathbf{q}}^{\mu}(\langle u_{\mathbf{k}}\vert u_{\mathbf{k}}\rangle)\left[\mathcal{D}_{\mathbf{q}}^{\nu}(\langle u_{\mathbf{k}}\vert u_{\mathbf{k}}\rangle)\right]^{*}\ell_{\mathbf{k},\mathbf{k}+\mathbf{q}}^{2}\frac{1}{2\Delta}.
\end{equation}
 In the limit $\mathbf{q}\rightarrow0$, although $p_{\mathbf{k},\mathbf{k}+\mathbf{q}}^{2}\rightarrow1$,
$\mathcal{D}_{\mathbf{q}}^{\mu}(\langle u_{\mathbf{k}}\vert u_{\mathbf{k}}\rangle)\rightarrow0$, thus
the response approaches zero,
\begin{equation}
R_{\mu\nu}(\mathbf{q}\rightarrow0,\omega=0)\rightarrow0.
\end{equation}
Therefore, the diamagnetic current tensor dominates the total current response tensor, implying a Messiner effect. Of course, in the normal phase,
the diamagnetic current vanishes as $\Delta=0$.

At finite temperatures, the quasiparticle state populations will be in
general non-zero. We have more contributions to the $R_{\mu\nu}$
\begin{align}
R_{\mu\nu}(\mathbf{q},\omega)= & \sum_{\mathbf{k}}\mathcal{D}_{\mathbf{q}}^{\mu}(\langle u_{\mathbf{k}}\vert u_{\mathbf{k}}\rangle)\left[\mathcal{D}_{\mathbf{q}}^{\nu}(\langle u_{\mathbf{k}}\vert u_{\mathbf{k}}\rangle)\right]^{*} \left\{-2\ell_{\mathbf{k},\mathbf{k}+\mathbf{q}}^{2}\frac{n_\mathrm{FD}(E_{\mathbf{k}+\mathbf{q}})-n_\mathrm{FD}(E_{\mathbf{k}})}{E_{\mathbf{k}}-E_{\mathbf{k}+\mathbf{q}}-\omega+i\eta}\right.\nonumber \\
 & \left.+p_{\mathbf{k},\mathbf{k}+\mathbf{q}}^{2}\left(\frac{n_\mathrm{FD}(E_{\mathbf{k}+\mathbf{q}})+n_\mathrm{FD}(E_{\mathbf{k}})-1}{\omega-(E_{\mathbf{k}}+E_{\mathbf{k}+\mathbf{q}})+i\eta}-\frac{n_\mathrm{FD}(E_{\mathbf{k}+\mathbf{q}})+n_\mathrm{FD}(E_{\mathbf{k}})-1}{\omega+(E_{\mathbf{k}}+E_{\mathbf{k}+\mathbf{q}})+i\eta}\right)\right\}.
\end{align}
When we take the limit $\omega=0,\mathbf{q}\rightarrow0$ where $\mathcal{D}_{\mathbf{q}}^{\mu}(\langle u_{\mathbf{k}}\vert u_{\mathbf{k}}\rangle)\rightarrow0$,
we still have 
\begin{equation}
R_{\mu\nu}(\mathbf{q}\rightarrow0,\omega=0)\rightarrow0.
\end{equation}
The paramagnetic current response vanishes even at finite temperatures, which contradicts the behavior observed in conventional superconductors. Consequently, there is no dissipation current in a flat-band superconductor. However, at finite temperatures, the pairing potential weakens, leading to a decrease in total demagnetization. Thus, the two-fluid model may not apply to a flat-band superconductor characterized by quantum metric.

\section{Example: Drude weight in flat band metal from interaction}

\begin{figure}
\includegraphics[scale=0.3]{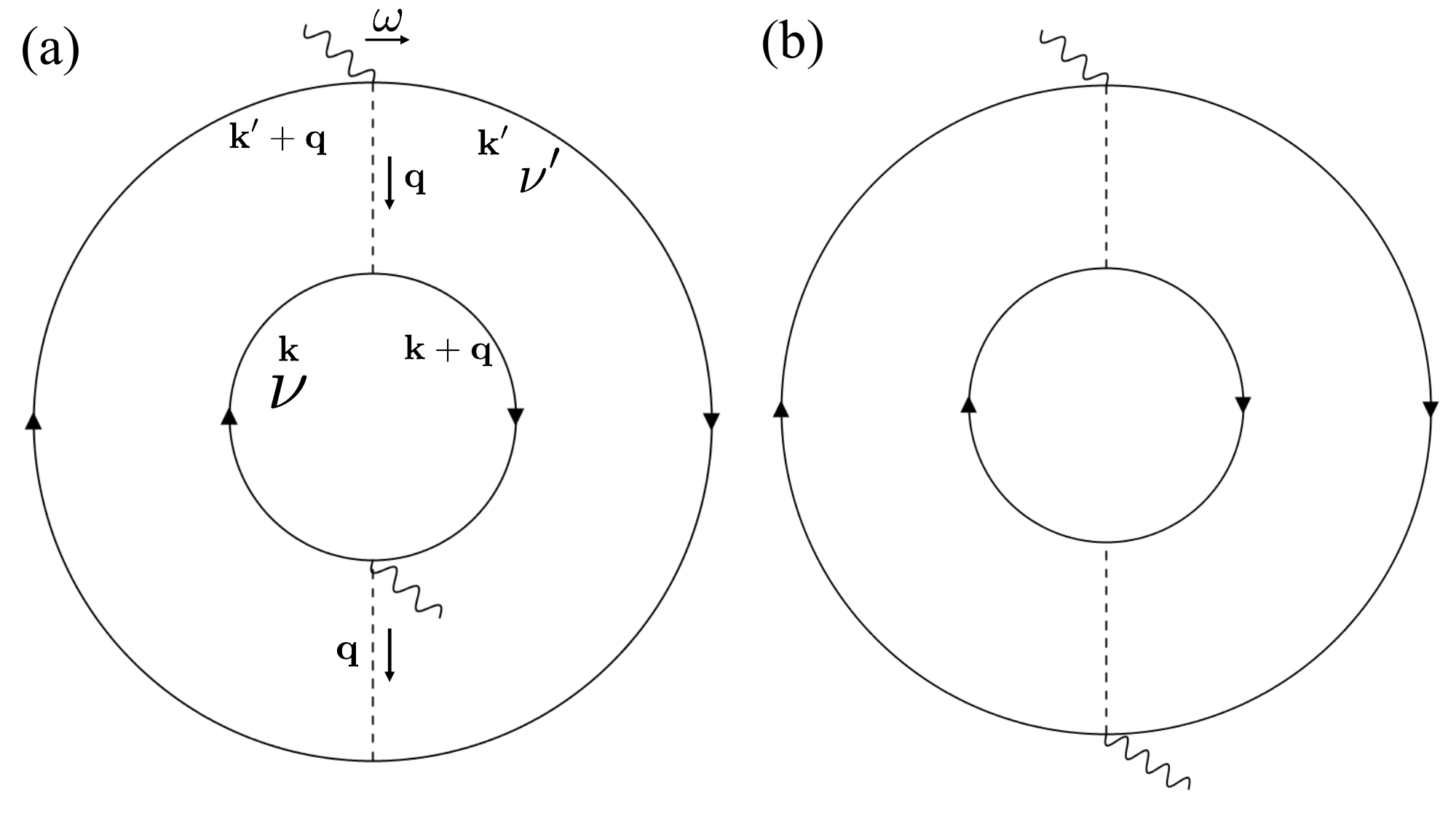}
\includegraphics[scale=0.315]{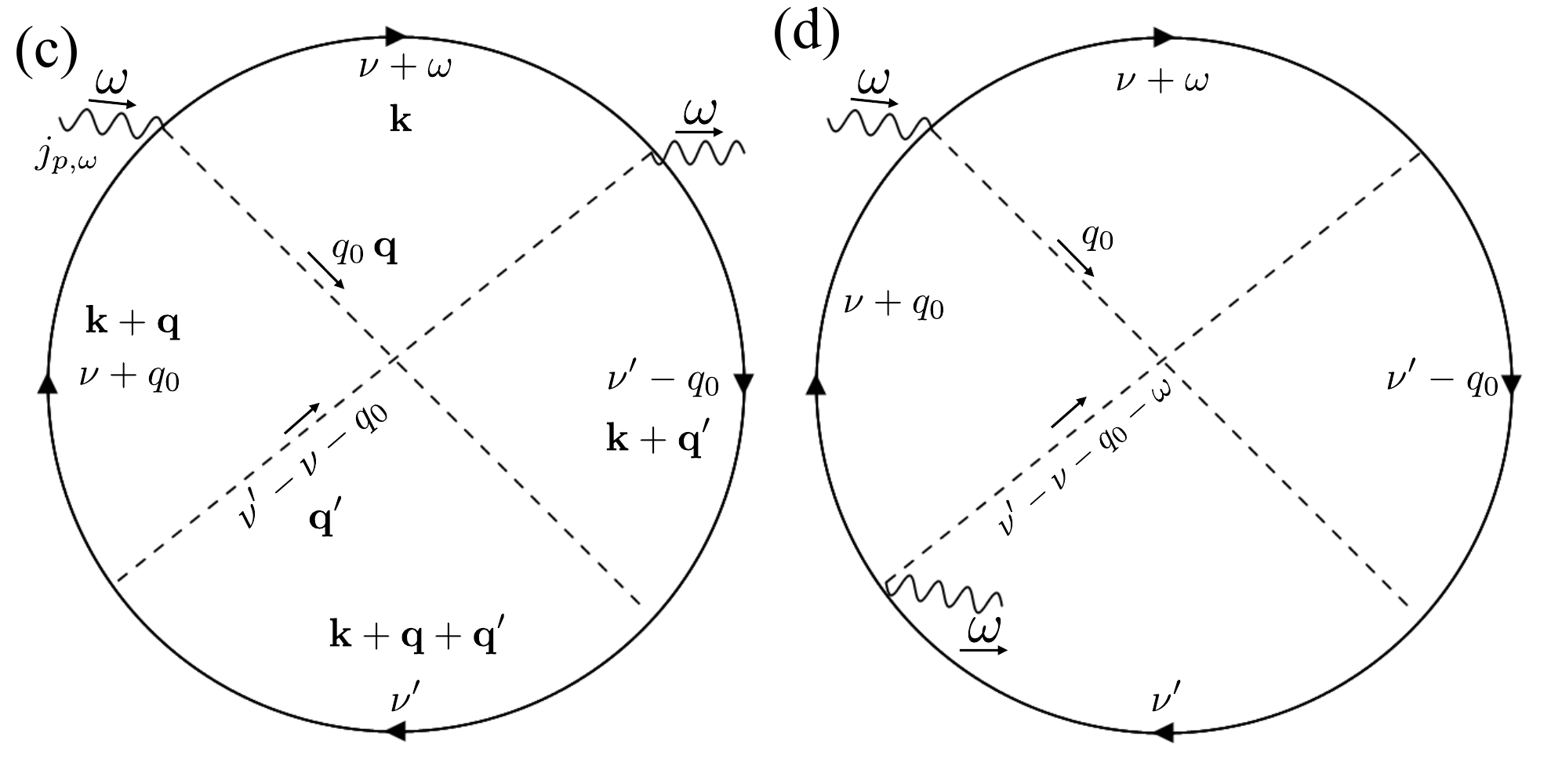}
\caption{Feynman diagrams of the Hartree contribution to the conductivity in the flat-band metal with the density-density interaction.}
\label{sm:hatree}
\end{figure}

The second example is to investigate the delocalization effect in
a flat-band metal.
The interaction we consider is the density-density
interaction,
\begin{equation}
H_{I}=\frac{1}{N}\sum_{\mathbf{k\mathbf{k}^{\prime}\mathbf{p}}}V_{\mathbf{p}}\rho_{\mathbf{k^{\prime}}}^{\dagger}(\mathbf{p})\rho_{\mathbf{k}}(\mathbf{p}),
\end{equation}
with $\rho_{\mathbf{k}}(\mathbf{p})=\sum_{\alpha}c_{\mathbf{k}\alpha}^{\dagger}c_{\mathbf{k}+\mathbf{p}\alpha}$.
The generalization Peierls substitution can lead to paramagnetic
current $j_{p,t}^{\mu}$
\begin{align}
j_{\mathrm{p},t}^{\mu} & =\frac{1}{N}\sum_{\mathbf{k}\mathbf{k}^{\prime}\mathbf{p}}V_{\mathbf{p}}\mathcal{J}_{\mathbf{k}\mathbf{k}^{\prime}}(\mathbf{p})\bar{\rho}_{\mathbf{k}^{\prime}}^{\dagger}(\mathbf{p})\bar{\rho}_{\mathbf{k}}(\mathbf{p}),\\
\mathcal{J}_{\mathbf{k}\mathbf{k}^{\prime}}^{\mu}(\mathbf{p}) & =\mathcal{D}_{\mathbf{0}}^{\mu}(\gamma_{\mathbf{k}^{\prime}}^{*}(\mathbf{p}))\gamma_{\mathbf{k}}(\mathbf{p})+\gamma_{\mathbf{k}^{\prime}}^{*}(\mathbf{p})\mathcal{D}_{\mathbf{0}}^{\mu}(\gamma_{\mathbf{k}}(\mathbf{p})).
\end{align}
with $\gamma_{\mathbf{k}}(\mathbf{p})=\langle u_{\mathbf{k}}\vert u_{\mathbf{k}+\mathbf{p}}\rangle$.

With the current operator, we can calculate the Drude weight which measures the charge stiffness. Within the linear response,
one calculates the response of the electric current $\langle j_\mathrm{p}^{\mu}\rangle$
to an electromagnetic gauge field $A_{\mathbf{q}}^{\nu}(t)$. For
the charge response of the normal state, one must always take the
limit $\mathbf{q}\rightarrow0$ before taking the frequency limit
$\omega\rightarrow0$. The Drude weight can be evaluated as 
\begin{align}
D_{\mu\nu} & =\pi\lim_{\omega+i\eta\rightarrow0}\left[\Pi^{\mu\nu}(iq_{0})-\Pi^{\mu\nu}(0)\right]_{iq_{0}\rightarrow\omega+i\eta},
\end{align}
with a small but positive $\eta$. Here 
 $\Pi^{\mu\nu}$ is the polarization
tensor
\begin{equation}
\Pi^{\mu\nu}=\langle j_{\mathrm{p}}^{\mu}(\omega)j_{\mathrm{p}}^{\nu}(-\omega)\rangle.
\end{equation}
This value is a Drude weight, which manifests in the conductivity
tensor as a delta function frequency response of the real part. Alternatively,
we can use another way to calculate the Drude weight 
\begin{equation}
D_{\mu\nu}=\lim_{\omega\rightarrow0}\omega\mathrm{Im}\Pi^{\mu\nu}(\omega),
\end{equation}
where it is easy to know that the diamagnetic current does not contribute
to the Drude weight. Equivalently, Drude weight measures the effective
electron density contributing to DC conductivity.

Back to the flat-band case, we can ignore the Fock term from the interaction when the temperature
is much larger than the bandwidth and the interaction strength. We
can apply the linear response theory to calculate the polarization
tensor 
\begin{equation}
\Pi^{\mu\nu}=\langle j_{\mathrm{p}}^{\mu}(\omega)j_{\mathrm{p}}^{\nu}(-\omega)\rangle,
\end{equation}
with $j_\mathrm{p}^\mu(\omega) = \int \frac{d\omega}{2\pi}j_{\mathrm{p},t}^\mu e^{i\omega t}$.
For regularization, we introduce a small dispersion which will be tuned
to zero finally. Fig.~\ref{sm:hatree} depicts the four types of the Hartree contributions. In Figs.~\ref{sm:hatree}(a) and (b), we have the contributions $\Pi_1^{\mu\nu}$,
\begin{align}
\Pi_{1}^{\mu\nu}(\omega)= & -2\int\frac{d\mathbf{k}d\mathbf{k}^{\prime}d\mathbf{p}}{(2\pi)^{6}}V_{\mathbf{p}}V_{-\mathbf{p}}\mathcal{J}_{\mathbf{k}\mathbf{k}}(\mathbf{p})\mathcal{J}_{\mathbf{k}^{\prime}\mathbf{k}^{\prime}}(-\mathbf{p})\mathcal{C}_{1}(\omega),
\end{align}
with 
\begin{align}
\mathcal{C}_{1}(\omega)= & T^{3}\sum_{\nu\nu^{\prime}q_{0}}\frac{1}{i\nu+iq_{0}-\epsilon_{\mathbf{q}}}\frac{1}{i\nu+i\omega-\epsilon_{\mathbf{k}+\mathbf{q}}}\frac{1}{i\nu^{\prime}-iq_{0}-\epsilon_{\mathbf{k}^{\prime}}}\frac{1}{i\nu^{\prime}-\epsilon_{\mathbf{k}^{\prime}+\mathbf{q}}}\nonumber \\
= & Tn_\mathrm{FD}^{\prime}(\epsilon)n_\mathrm{FD}^{\prime}(\epsilon)+T^{3}\sum_{q_{0}\neq0}\sum_{\nu\nu^{\prime}}\frac{n_\mathrm{FD}(\epsilon_{1})-n_\mathrm{FD}(\epsilon_{2})}{iq_{0}-\epsilon_{1}+\epsilon_{2}}\frac{1}{i\nu+i\omega-\epsilon_{2}}\frac{1}{i\nu^{\prime}-iq_{0}-\epsilon_{3}}\frac{1}{i\nu^{\prime}-\epsilon_{4}}.
\end{align}
In Figs.~\ref{sm:hatree}(c) and (d), the Hartree diagram has the expression, 
\begin{align}
\Pi_{2}^{\mu\nu}= & \int\frac{d\mathbf{k}d\mathbf{p}^{\prime}d\mathbf{p}}{(2\pi)^{6}}V_{\mathbf{p}}V_{-\mathbf{p}^{\prime}}\mathcal{J}_{\mathbf{k}\mathbf{k}}(\mathbf{p})\mathcal{J}_{\mathbf{\mathbf{k}k}}(-\mathbf{p}^{\prime})\mathcal{C}_{2}(\omega),
\end{align}
with 
\begin{align}
\mathcal{C}_{2}(\omega) & =T^{3}\sum_{\nu\nu^{\prime}q_{0}}\frac{1}{i\nu+iq_{0}-\epsilon_{\mathbf{k}+\mathbf{q}}}\frac{1}{i\nu+i\omega-\epsilon_{\mathbf{k}}}\frac{1}{i\nu^{\prime}-iq_{0}-\epsilon_{\mathbf{k}+\mathbf{q}^{\prime}}}\frac{1}{i\nu^{\prime}-\epsilon_{\mathbf{k}+\mathbf{q}+\mathbf{q}^{\prime}}}\nonumber \\
 & =Tn_\mathrm{FD}^{\prime}(\epsilon)n_\mathrm{FD}^{\prime}(\epsilon)+T^{3}\sum_{q_{0}\neq0}\sum_{\nu\nu^{\prime}}\frac{1}{i\nu+iq_{0}-\epsilon_{\mathbf{k}+\mathbf{q}}}\frac{1}{i\nu+i\omega-\epsilon_{\mathbf{k}}}\frac{1}{i\nu^{\prime}-iq_{0}-\epsilon_{\mathbf{k}+\mathbf{q}^{\prime}}}\frac{1}{i\nu^{\prime}-\epsilon_{\mathbf{k}+\mathbf{q}+\mathbf{q}^{\prime}}}.
\end{align}
Therefore, we have the Drude weight
\begin{equation}
D_{\mu\nu}(T)=\frac{\pi(\bar\nu(1-\bar\nu))^{2}}{T}\left[\int\frac{d\mathbf{k}d\mathbf{k}^{\prime}d\mathbf{p}}{(2\pi)^{6}}V_{\mathbf{p}}V_{-\mathbf{p}}\mathcal{J}_{\mathbf{k}\mathbf{k}}(\mathbf{p})\mathcal{J}_{\mathbf{k}^{\prime}\mathbf{k}^{\prime}}(-\mathbf{p})-\int\frac{d\mathbf{k}d\mathbf{p}^{\prime}d\mathbf{p}}{(2\pi)^{6}}V_{\mathbf{p}}V_{-\mathbf{p}^{\prime}}\mathcal{J}_{\mathbf{k}\mathbf{k}}(\mathbf{p})\mathcal{J}_{\mathbf{\mathbf{k}k}}(-\mathbf{p}^{\prime})\right],
\end{equation}
with $\bar\nu$ being the filling factor.

\section{Example: Delocalization by disorder in a flat system}
The third example in the main text is the disorder in a flat band.
We consider a multi-band system with Hamiltonian 
\begin{align}
H_{0} & =\sum_{\mathbf{k}}\sum_{\alpha\beta}h_{\alpha\beta}(\mathbf{k})c_{\mathbf{k}\alpha}^{\dagger}c_{\mathbf{k}\beta},
\end{align}
which possesses an isolated ideal flat band around the Fermi energy.
The $H_{0}$ can be diagonalized with the Bloch waves $\vert u_{n}(\mathbf{k})\rangle$:
$H\vert u_{n}(\mathbf k)\rangle=E_{n}(\mathbf k)\vert u_{n}(\mathbf k)\rangle$. We can project
the system to the flat band through $c_{\mathbf k\alpha}\rightarrow u_{\mathbf k\alpha}^{*}f_{\mathbf k}$.
We realize an on-site disorder with 
\begin{equation}
H_{I}=\sum_{\mathbf{r}\alpha}w_{\mathbf{r}}c_{\mathbf{r}\alpha}^{\dagger}c_{\mathbf{r}\alpha}.
\end{equation}
For simplicity, we require the disorder average 
\begin{equation}
\overline{w_{\mathbf{r}}w_{\mathbf{r}^{\prime}}}=\gamma^{2}\delta_{\mathbf{r}\mathbf{r}^{\prime}},
\end{equation}
where $\overline{\cdot}$ denotes the disorder averaging. 

In the following, we focus on the two-dimensional case. We point out
that we do not intend to give a complete theory on the disorder effect
of a flat-band system. Our purpose is merely to show the current and
the possible conductivity in the leading order. 

\subsection{Current operator by disorder }

We can derive the current operator in the subspace $\mathbb{H}^{\text{sub}}$.
We can introduce the Lagrange multipliers $\lambda_{\mathbf{k}\alpha}$
and $\lambda_{\mathbf{k}\alpha}^{\dagger}$ to realize the band projection
\begin{equation}
L_{\lambda}=\sum_{\mathbf{k}}\lambda_{\mathbf{k}\alpha}^{\dagger}(c_{\mathbf{k}\alpha}-u_{\mathbf{k}\alpha}^{*}f_{\mathbf{k}})+h.c.,
\end{equation}
We have the total Lagrange 
\begin{equation}
L=\sum_{\mathbf{k\alpha}}c_{\mathbf{k}\alpha}^{\dagger}i\partial_{t}c_{\mathbf{k}\alpha}-H_{0}-H_{I}-L_{\lambda},
\end{equation}
We need to apply the Peierls substitution to both $H_{0}$ and the
$L_{\lambda}$.We consider $A_{t}^{\mu}$ which is the time-dependent
gauge potential to realize the static electromagnetic field in the
limit 
\begin{equation}
A_{\omega}^{a}=E(\omega)\left[\frac{1}{i\omega}-\pi\delta(\omega)\right].
\end{equation}
For the Lagrange multiplier term $L_{\lambda}$, we have 
\begin{align}
L_{\lambda}[A] & =\sum_{\mathbf{k}\alpha}[\lambda_{\mathbf{k}\alpha}^{\dagger}(\nu)(c_{\mathbf{k}\alpha}(\nu)-u_{\mathbf{k}\alpha}^{*}f_{\mathbf{k}}(\nu))+\sum_{\mu}A_{\omega}^{\mu}\mathcal{D}_{\mu}u_{\mathbf{k}\alpha}^{*}\lambda_{\mathbf{k}\alpha}^{\dagger}(\nu)f_{\mathbf{k}}(\nu+\omega)\nonumber \\
 & -\sum_{\mu\nu}\frac{1}{2}A_{\omega}^{\mu}A_{-\omega}^{\nu}\mathcal{D}_{\mu}\mathcal{D}_{\nu}u_{\mathbf{k}\alpha}^{*}\lambda_{\mathbf{k}\alpha}^{\dagger}(\nu)f_{k}(\nu)]+h.c.,
\end{align}
with $\nu=2\pi(n+1)T$ is the Matsubara frequency. It is easy to check that the constraint 
becomes 
\begin{align}
c_{\mathbf{k}\alpha}(\nu) & =u_{\mathbf{k}\alpha}^{*}f_{\mathbf{k}}(\nu)-\sum_\mu A_{\omega}^{\mu}\mathcal{D}_{\mu}u_{\mathbf{k}\alpha}^{*}f_{\mathbf{k}}(\nu+\omega) -\sum_{\mu\nu}\frac{1}{2} A_{\omega}^{\mu}A_{-\omega}^{\nu}G_{\mu\nu}(\mathbf{k})u_{\mathbf{k}\alpha}^{*}f_{\mathbf{k}}(\nu).\label{eq:ckA}
\end{align}

Substitute Eq.~(\ref{eq:ckA}) to the disorder term $H_{\text{int}}$,
we have the current operator 
\begin{align}
j_{\omega}^{\mu} & =T\sum_\nu\sum_{\mathbf{p}}w_{\mathbf{p}}\left[\langle u_{\mathbf{k+\mathbf{p}}}\vert\mathcal{D}_{\mu}u_{\mathbf{k}}\rangle+\langle\mathcal{D}_{\mu}u_{\mathbf{k+\mathbf{p}}}\vert u_{\mathbf{k}}\rangle\right]f_{\mathbf{k}}^{\dagger}(\nu)f_{\mathbf{k}+\mathbf{p}}(\nu+\omega)\nonumber \\
 & \equiv T\sum_\nu \sum_{\mathbf{p}}w_{\mathbf{p}}\mathcal{J}_{\mathbf{k}}^{\mu}(\mathbf{p})f_{\mathbf{k}}^{\dagger}(\nu)f_{\mathbf{k}+\mathbf{p}}(\nu+\omega),
\end{align}
Here $w_{\mathbf{p}}=1/N\sum_{\mathbf{r}}w_{\mathbf{p}}e^{i\mathbf{p}\cdot\mathbf{r}}$. 

In summary, we have the partition function 
\begin{equation}
Z[A]=\overline{\int Df_{\mathbf{k}\alpha}Df_{\mathbf{k}\alpha}^{\dagger}\exp\left(i\int dtL[f,A]\right)},
\end{equation}
with the Lagrange
\begin{equation}
L[f,A]=\sum_{\mathbf{k}}if_{\mathbf{k}}^{\dagger}\partial_{t}f_{\mathbf{k}}-\sum_{\mathbf{k}\mathbf{p}}w_{\mathbf{p}}\langle u_{\mathbf{k+\mathbf{p}}}\vert u_{\mathbf{p}}\rangle f_{\mathbf{k}}^{\dagger}f_{\mathbf{k}+\mathbf{p}}-\sum_{\mu}A_{\omega}^{\mu}j_{\omega}^{\mu}.\label{smeq:Leff}
\end{equation}

\subsection{Single-particle Green function and Dyson equation }

We evaluate the single-particle Green function renormalized by disorder.
Without disorder, we have the retarded Green function 
\begin{equation}
G_{0}^{R}(\omega,\mathbf{k})=i\langle f_{\mathbf{k}}(\omega)f_{\mathbf{k}}^{\dagger}(\omega)\rangle=\frac{1}{\omega-i\eta}.
\end{equation}
The disorder averaged Green function $\overline{G^{R}(\omega,\mathbf{k})}$
can be obtained by the self-consistent Born approximation. With the
effective action, the Dyson equation can be
formulated as 
\begin{equation}
\overline{G^{R}(\omega,\mathbf{k})}=G_{0}^{R}(\omega,\mathbf{k})+G_{0}^{R}(\omega,\mathbf{k})\Sigma(\omega,k)\overline{G^{R}(\omega,\mathbf{k})} ,
\end{equation}
with 
\begin{equation}
\Sigma(\omega,k)=\gamma^{2}\int\frac{d\mathbf{q}}{(2\pi)^{2}}\vert\langle u_{\mathbf{k}+\mathbf{q}}\vert u_{\mathbf{k}}\rangle\vert^{2}\overline{G^{R}(\omega,\mathbf{k}+\mathbf{q})}.
\end{equation}
At leading order we can ignore the band dispersion induced by
the disorder as we are interested in a flat band limit. The dispersion
effect may be investigated independently by the Hartree-Fock approximation
the disorders. Therefore, we can approximate 
\begin{equation}
\Sigma(\omega,k)=\gamma^{2}\int\frac{d\mathbf{q}}{(2\pi)^{2}}\overline{G^{R}(\omega,\mathbf{k}+\mathbf{q})}.
\end{equation}
We assume the solution to be independent of the momentum 
\begin{equation}
\overline{G^{R}(\omega,\mathbf{k})}\equiv\overline{G^{R}(\omega)}=\frac{1}{\omega-\epsilon-i\Sigma(\omega)},
\label{eq:greenR}
\end{equation}
which yields a self-consistent equation
\begin{equation}
\Sigma(\omega)=\gamma^{2}\frac{1}{\omega-\epsilon-i\Sigma(\omega)}.
\end{equation}
We can find a solution 
\begin{equation}
\Sigma(\omega)=\begin{cases}
\frac{(\omega-\epsilon)-\sqrt{(\omega-\epsilon)^{2}-4\gamma^{2}}}{2}  & \omega>2\gamma\\
\frac{(\omega-\epsilon)+i\sqrt{4\gamma^{2}-(\omega-\epsilon)^{2}}}{2} & \vert\omega\vert<2\gamma\\
\frac{(\omega-\epsilon)+\sqrt{(\omega-\epsilon)^{2}-4\gamma^{2}}}{2}  & \omega<-2\gamma
\end{cases}
\end{equation}
Notice that there should be no disorder effect when the $\omega$
is greatly larger than $\gamma$. Namely, we have the tendency $\overline{G^{R}(\omega)}=\frac{1}{\omega-i0^{+}}$
for $\omega\gg\vert\gamma\vert$. This condition helps fix the solution
of the self-energy $\Sigma$. In particular for small $\omega$, we
have 
\begin{align}
\Sigma(\omega) & =\frac{\omega}{2}+i\gamma,\\
\overline{G^{R}(\omega,\mathbf{k})} & =\frac{1}{\frac{\omega}{2}+i\gamma}.
\end{align}
One may find that the bands now get broadened by the disorder effect.
We can evaluate the density of states
\begin{equation}
\rho(\omega)=-\frac{1}{\pi}\int\frac{d\mathbf{k}}{(2\pi)^{2}}\mathrm{Im}\overline{G^{R}(\omega,\mathbf{k})}=\begin{cases}
\frac{1}{\pi}\frac{\sqrt{4\gamma^{2}-\omega^{2}}}{2\gamma^{2}} & \vert\omega\vert<2\gamma\\
0 & \vert\omega\vert>2\gamma
\end{cases}
\end{equation}
which is consistent with the naive picture.

\subsection{Current-current correlators and conductivity}

With the effective Lagrangian in Eq.~\eqref{smeq:Leff}, we can calculate the polarization tensor, that is, the current-current correlator,
\begin{align}
\overline{\langle j_{\omega}^{\mu}j_{-\omega}^{\nu}\rangle} & =iT^2\sum_{\nu_n\nu_n^\prime}\int\frac{d\mathbf{q}d\mathbf{q}^{\prime}}{(2\pi)^{4}}\int\frac{d\mathbf{k}d\mathbf{k}^{\prime}}{(2\pi)^{4}}[n_\mathrm{FD}(\nu_n)-n_\mathrm{FD}(\nu_n+\omega)]\langle w_{\mathbf{p}}w_{-\mathbf{p}}\mathcal{J}_{\mathbf{k}}^{\mu}(\mathbf{p})\mathcal{J}_{\mathbf{k}^{\prime}}^{\nu}(-\mathbf{p})\overline{\langle f_{\mathbf{k}}^{\dagger}(\nu_n)f_{\mathbf{k}+\mathbf{p}}(\nu_n+\omega)f_{\mathbf{k}^{\prime}+\mathbf{p}}^{\dagger}(\nu_n^{\prime})f_{\mathbf{k}^{\prime}}(\nu_n^{\prime}-\omega)\rangle}\nonumber \\
 & =iT^2\sum_{\nu_n\nu_n^\prime}\gamma^{2}\int\frac{d\mathbf{q}}{(2\pi)^{2}}\int\frac{d\mathbf{k}d\mathbf{k}^{\prime}}{(2\pi)^{4}}[n_\mathrm{FD}(\nu_n)-n_\mathrm{FD}(\nu_n+\omega)]\mathcal{J}_{\mathbf{k}}^{\mu}(\mathbf{p})\mathcal{J}_{\mathbf{k^{\prime}}}^{\nu}(-\mathbf{p})\overline{\langle f_{\mathbf{k}}^{\dagger}(\nu_n)f_{\mathbf{k}+\mathbf{p}}(\nu_n+\omega)f_{\mathbf{k}^{\prime}+\mathbf{p}}^{\dagger}(\nu_n^{\prime})f_{\mathbf{k}^{\prime}}(\nu_n^{\prime}-\omega)\rangle}\nonumber \\
 & \equiv i\gamma^{2}\int\frac{d\mathbf{q}} {(2\pi)^{2}}\int\frac{d\mathbf{k}d\mathbf{k}^{\prime}}{(2\pi)^{4}}\mathcal{J}_{\mathbf{k}}^{\mu}(\mathbf{p})\mathcal{J}_{\mathbf{k}}^{\nu*}(\mathbf{p})\mathcal{C}(\mathbf{p},\omega),
\end{align}
with 
\begin{equation}
\mathcal{C}(\mathbf{p},\omega)=T^2\sum_{\nu\nu^\prime}\int\frac{d\mathbf{k}d\mathbf{k}^{\prime}}{(2\pi)^{4}}[n_\mathrm{FD}(\nu)-n_\mathrm{FD}(\nu+\omega)]\overline{\langle f_{\mathbf{k}}^{\dagger}(\nu)f_{\mathbf{k}+\mathbf{p}}(\nu+\omega)f_{\mathbf{k}^{\prime}+\mathbf{p}}^{\dagger}(\nu^{\prime})f_{\mathbf{k}^{\prime}}(\nu^{\prime}-\omega)\rangle}.
\end{equation}
For the scope of this paper, we can keep the zeroth order without
regarding the quantum corrections like diffuson and Cooperon contributions,
\begin{align}
\mathcal{C}(\mathbf{q},\omega) & =\gamma^{2}\int\frac{d\mathbf{k}d\mathbf{k}^{\prime}}{(2\pi)^{4}}\int_{-2\gamma}^{2\gamma}\frac{d\nu}{2\pi}\mathcal{J}_{\mathbf{k}}^{\mu}(\mathbf{p})\mathcal{J}_{\mathbf{k}}^{\nu*}(\mathbf{p})[n_\mathrm{FD}(\nu)-n_\mathrm{FD}(\nu+\omega)]\overline{G^{R}(\nu,k)}\overline{G^{A}(\nu+\omega,k+q)}\delta(\mathbf{k}-\mathbf{k}^{\prime})\nonumber \\
 & =\frac{\gamma^{2}}{(2\pi)^{2}}\int\frac{d\mathbf{k}}{(2\pi)^{2}}\mathcal{J}_{\mathbf{k}}^{\mu}(\mathbf{p})\mathcal{J}_{\mathbf{k}}^{\nu*}(\mathbf{p})[n_\mathrm{FD}(\nu)-n_\mathrm{FD}(\nu+\omega)]\frac{2}{\nu-i\sqrt{4\gamma^{2}-\nu^{2}}}\frac{2}{\omega+\nu+i\sqrt{4\gamma^{2}-(\omega+\nu)^{2}}}\nonumber \\
 & =\frac{\gamma^{2}}{(2\pi)^{2}}\int\frac{d\mathbf{k}}{(2\pi)^{2}}\mathcal{J}_{\mathbf{k}}^{\mu}(\mathbf{p})\mathcal{J}_{\mathbf{k}}^{\nu*}(\mathbf{p})\frac{2}{-i2\gamma}\frac{2\omega}{\omega+i\sqrt{4\gamma^{2}-\omega{}^{2}}},
\end{align}
which gives rise to 
\begin{align}
\overline{\langle j_{\omega}^{\mu}j_{-\omega}^{\nu}\rangle} & =\gamma^{2}\int\frac{d\mathbf{q}d\mathbf{k}}{(2\pi)^{2}}\mathcal{J}_{\mathbf{k}}^{\mu}(\mathbf{p})\mathcal{J}_{\mathbf{k}}^{\nu*}(\mathbf{p})\frac{i\omega}{\gamma^{2}}.
\end{align}
Therefore we have the DC conductivity,
\begin{align}
\sigma_{\mu\nu} & =\lim_{\omega\rightarrow0}\frac{\Pi^{\mu\nu}(\omega)-\Pi^{\mu\nu}(0)}{i\omega}\nonumber \\
 & =\frac{1}{(2\pi)^{2}}\int\frac{d\mathbf{k}d\mathbf{p}}{(2\pi)^{4}}\mathcal{J}_{\mathbf{k}}^{\mu}(\mathbf{p})\mathcal{J}_{\mathbf{k}}^{\nu*}(\mathbf{p}).
\end{align}
In two dimensions, conductivity is dimensionless. The derived conductivity indicates potential transport influenced by quantum geometry. However, this result should not be viewed as a comprehensive theory. It falls short of adequately describing a two-dimensional disordered system, where disorder plays a crucial role in driving an Anderson insulator.

\end{document}